\documentclass[twocolumn,preprintnumbers,amsmath,amssymb,]{revtex4}

\usepackage{graphicx}
\usepackage{dcolumn}
\usepackage{bm}
\usepackage{xcolor}
\usepackage{subfigure}
\usepackage{dcolumn}
\usepackage{txfonts}
\usepackage{multirow}
\usepackage{subfloat}
\usepackage[colorlinks,linkcolor=blue,anchorcolor=blue,citecolor=blue]{hyperref}
\usepackage{blindtext}
\usepackage{comment}
\usepackage{times}

\newcommand{\Dcal}{\mathcal{D}}

\newcommand{\mb}[1]{\boldsymbol{#1}}
\newcommand{\br}{\mb{r}}

\newcommand{\DM}[2]{| {#1} \rangle\langle{#2} |}

\newlength{\back}

\begin{document}

\title{Chemical potential, derivative discontinuity, fractional electrons, jump of the Kohn-Sham potential, atoms as thermodynamic open systems, and other (mis)conceptions of the density functional theory of electrons in molecules}
\author{E. J. Baerends}
\affiliation{Vrije Universiteit, Amsterdam, The Netherlands}

\date{\today}

\begin{abstract}
\noindent
Many references exist in the density functional theory (DFT) literature to the chemical potential of the electrons in an atom or a molecule. The origin of this notion has been the identification of the Lagrange multiplier $\mu = \partial E/\partial N$ in the Euler-Lagrange variational equation for the ground state density as the chemical potential of the electrons.  We first discuss why the Lagrange multiplier in this case is an arbitrary constant and therefore cannot be a physical characteristic of an atom or molecule. The switching of the energy derivative (``chemical potential'') from $-I$ to $-A$ when the electron number crosses the integer, called integer discontinuity or derivative discontinuity, is not physical but only occurs when the nonphysical noninteger electron systems and the corresponding energy and derivative $\partial E/\partial N$ are chosen in a specific discontinuous way. The question is discussed whether in fact the thermodynamical concept of a chemical potential can be defined for the electrons in such few-electron systems as atoms and molecules. The conclusion is that such systems lack important characteristics of thermodynamic systems and do not afford the definition of a chemical potential. They also cannot be considered as analogues of the open systems of thermodynamics that can exchange particles with an environment (a particles bath or other members of a Gibbsian ensemble).  Thermodynamical (statistical mechanical) concepts like chemical potential, open systems, grand canonical ensemble etc. are not applicable to a few electron system like an atom or molecule. A number of topics in DFT are critically reviewed in the light of these findings: jumps in the Kohn-Sham potential when crossing an integer number of electrons, the band gap problem, the deviation-from-straight-lines error, the role of ensembles in DFT.
\end{abstract}

\maketitle

\section{Introduction}\label{sec:introduction}
Hohenberg and Kohn \cite{HohenbergKohn1964} have established the unique correspondence between the ground state density of an $N$-electron system in an external local potential and its ground state wavefunction and electron density. This implies that the ground state energy is a functional of the $N$-electron ground state density. They also derived for the corresponding functional $E_v[\rho]$ the variational property of being a minimum on the domain of ground state densities. Leaving aside questions of definition of the functional on appropriate density domains, we note that this has led to the formulation of the Euler-Lagrange variational equation for the determination of the ground state density:
\begin{equation}\label{eq:Euler-Lagrange}
\frac{\delta }{\delta \rho(\br)} \left[ E_v[\rho] -\mu\left(\int \rho(\br)d\br - N\right)\right] =0
\end{equation}
($N$ is integer).  The constraint that  the total number of electrons must be the integer $N$ is built in with the Lagrange multiplier $\mu$.  According to the theory of Lagrange multipliers one should have $\mu = \partial E_v / \partial N$. This looks like the chemical potential of thermodynamics and $\mu$ has indeed been called \cite{ParrDonnelly1978} ``a characteristic of the system of interest to be denoted the chemical potential of the (electrons of the) system''. This has been widely quoted and is often considered an important feature of the fundamentals of DFT \cite{Geerlings2003ChemRev,Geerlings2021}.  However, a careful consideration of the application of the integer-electron constraint  with the Lagrange multiplier technique does not support the attribution of physical meaning to the Lagrange multiplier in this case \cite{Baerends2020Janak}. The problem is in the definition of $\partial E_v/ \partial N$. The natural definition is \cite{DreizlerGross1990}
\begin{align}\label{eq:partialEpartialN}
\frac{\partial E_v}{\partial N}= \lim_{\epsilon \to 0} \frac{E_v[\rho_{N+\epsilon}]-E_v[\rho_N]}{\epsilon}
\end{align}
But what is $E_v[\rho_{N+\epsilon}]$? The HK theorem has been explicitly derived for $N$-electron ($N$ integer) systems. The HK theorem is rooted in quantum mechanics, using the wavefunctions of $N$-electron systems. There is no wavefunction for an $(N+\epsilon)$-electron system. Such a system does not exist. So $E_v[\rho_{N+\epsilon}]$ does not exist. That is why we have to restrict the variations in $\rho$ to $N$-conserving ones in the first place. \\
The restricted domain of densities on which $E_v[\rho]$ is defined is called in optimization theory the domain of feasible densities, or the feasible domain for short.  Constrained derivatives, which only consider infinitesimal variations of the variable (i.e.\ the function in functional analysis) over the feasible domain, are difficult to handle. The crux of the Lagrange multiplier method is that it allows one to use full derivatives. This requires that the full derivative is defined, including its components. But we have just noted that the crucial component $\partial E_v / \partial N$ is not defined. We will discuss this problem in section \ref{sec:Euler-Lagrange}, cf.\ Ref.~\cite{Baerends2020Janak}. From this discussion it emerges that the Lagrange multiplier in \eqref{eq:Euler-Lagrange} is an arbitrary constant. It does not have physical meaning and cannot be interpreted as the chemical potential of the electrons in an atom or molecule. The DFT literature is nevertheless replete with references to this ``chemical potential''. Next we will argue, in section \ref{sec:PPLB} that in fact the meaning of a chemical potential for the few electrons in an atom or molecule is problematic. Against the background of a summary of the well known statistical mechanical underpinning of thermodynamics  in appendix \ref{app:statistics}, it is demonstrated that the thermodynamic origin of the concept of chemical potential is not compatible with a few-electron quantum mechanical system that does not obey the characteristic properties of a macroscopic thermodynamic system. \\
\\
In a well-know paper Perdew et al.\ \cite{PerdewParrLevyBalduz1982} (PPLB) have highlighted the paradox that arises when $\mu$ of Eq.~\eqref{eq:Euler-Lagrange} is considered a chemical potential. As a solution they propose to extend the domain of densities on which $E_v[\rho]$ is  defined by introducing an ensemble of  two states with different electron numbers. It has been argued (see Ref.\ \cite{Baerends2020Janak} and section \ref{sec:Euler-Lagrange}) that this does not solve the problems with the identification of $\mu$ in \eqref{eq:Euler-Lagrange} as a chemical potential. \\
\\
We note in passing that similar criticism as the one here against the quantity $\partial E_v[\rho] / \partial N$ can be levelled against the derivative $\partial E_v[\rho] / \partial n_i$, where $n_i$ is the occupation number of orbital $\phi_i$ in the Kohn-Sham approach of DFT. In KS DFT the equality $\partial E_v[\rho] / \partial n_i=\epsilon_i$ is usually assumed and denoted Janak's theorem \cite{Janak1978}. Again the derivative is defined as
\begin{align}\label{eq:partialEpartialn_i}
\frac{\partial E_v}{\partial n_i}= \lim_{\delta \to 0} \frac{E_v[\rho(n_i+\delta)]-E_v[\rho^N]}{\delta}
\end{align}
But what is $E_v[\rho(n_i+\delta)]$? The Kohn-Sham system of noninteracting electrons has $N$ particles, which each occupy a one-electron wavefunction (spin orbital). It is not even clear what it means to say that orbital $\phi_i$ is occupied by $n_i+\delta$ electrons: this is a nonexisting system for which the energy or (KS) wavefunction cannot be known. More detailed discussion of the problems with Janak's theorem is given in Ref.~\cite{Baerends2020Janak}. We note that approximate expressions for the energy can be given, such as Hartree-Fock or exchange-only LDA (X$\alpha$) or GGAs, in which occupation numbers can be introduced. That implies that such an energy, although not physical, is mathematically defined at noninteger electron number, and derivatives can be taken \cite{Baerends2018JCP}. This has originally been introduced by Slater \cite{SlaterMann1969,Slater1972,Slater1974Vol4} and used in his transition state method for ionization and excitation energies. The old relation $\partial E^{appr}[\rho] / \partial n_i=\epsilon_i^{appr}$ can be called Slater's relation.  It is only applicable if in the approximate energy expression occupation numbers have been introduced in such a way that this relation can be derived \cite{Baerends2018JCP}. That can be done for approximations like the Hartree-Fock model, for X$\alpha$ and LDA and in (semi)-local and hybrid DFAs. \\
\\
The structure of this article is as follows. In section \ref{sec:Euler-Lagrange}  the arbitrariness of the Lagrange multiplier $\mu=\partial E_v/ \partial N$ is discussed. This raises the question of the validity of the concept of chemical potential for the electrons in an atom or molecule. In the following section (section \ref{sec:PPLB}), it is argued that indeed electrons in an atom or molecule do not have the properties that would allow to treat them as a thermodynamic system to which the laws of statistics (arising from the exceedingly large numbers of particles that feature in thermodynamic systems) and thermodynamic concepts such as chemical potential and temperature would be applicable. Section \ref{sec:steps} deals with the conditions and conclusions for the step behavior of functions like $\mu(\overline{N})$ and $\overline{E}(\overline{N})$ which follow from the PPLB Ansatz for the grand canonical ensemble-like probability distribution of neutral atom and positive and negative ion. The findings in sections \ref{sec:Euler-Lagrange} - \ref{sec:steps} have a bearing on several topics that feature frequently in DFT. These are touched upon in section \ref{sec:Miscellaneous}:  Steps in the KS potential in section \ref{subsec:potentialsteps}, the band gap problem of solid state physics in section \ref{subsec:BandGap}, the issue of atoms as open systems with a fluctuating electron number in section \ref{subsec:fluctuation}, the straight-lines condition in section \ref{subsec:DLSE} and the use of ensembles in DFT in section \ref{subsec:GOK}.  Section \ref{sec:Conclusions} makes summarizing remarks.\\
In Appendix A a brief review is given of the statistical mechanical underpinning of thermodynamics.  Although unabashedly unoriginal, we need this exposition to establish the salient features of statistical mechanics which prevent the treatment of few-electron quantum mechanical systems (atoms and molecules) as thermodynamic systems. It can be skipped by anyone familiar with statistical mechanics and thermodynamics. Appendix B discusses how  cases should be understood where properties like chemical potential  and temperature are attributed to (particles in) small subsystems of macroscopic thermodynamic systems. \\
Ref.\ \cite{Baerends2020Janak} dealt with the elucidation of the derivatives \eqref{eq:partialEpartialN} and \eqref{eq:partialEpartialn_i} and the consequences. It was concerned with the $T=0$ situation exclusively. The present paper replaces and corrects statements in Ref.\ \cite{Baerends2020Janak} referring to the finite temperature situation and its statistical mechanical treatment.

\section{The chemical potential interpretation of the Lagrange multiplier in the Euler-Lagrange equation (1) of DFT} \label{sec:Euler-Lagrange} 
    
We write the density as product of a shape factor $\sigma(\br)$ times the number of electrons \cite{ParrBartolotti1983}, the shape factor integrating to 1:
\begin{align}\label{eq:Nsigma}
\rho(\br) = N\sigma(\br), \quad \sigma(r) \equiv \rho(\br)/N, \quad \int \sigma(\br)d\br =1
\end{align} 
In the present case one distinguishes as  partial derivatives the one with respect to $N$, while keeping $\sigma$ constant, and the partial derivative with respect to density shape but with constant number of electrons,
\begin{align} \label{eq:deltaE[rho]}
\frac{\delta E_v[\rho]}{\delta \rho(\br)} &=  \left(\frac{\delta E_v[\rho]}{\delta \rho(\br)}\right)_{\sigma(\br)}+\left(\frac{\delta E_v[\rho]}{\delta \rho(\br)} \right)_N \notag \\
&\equiv \frac{\partial E_v[\rho]}{\partial N}+ \frac{\delta E_v[\rho]}{\delta \sigma(\br)}
\end{align}
While with the HK theorem the partial derivative with respect to density shape, $(\delta E_v[\rho]/\delta \sigma(\br))_{N}$, is defined, this is not the case for the derivative with respect to $N$, $(\partial E_v[\rho] /\partial N)_{\sigma(\br)}$, see Eq.\ \eqref{eq:partialEpartialN}. (In analogy to partial derivatives of functions of more variables $(\partial E_v[\rho] /\partial N)_{\sigma(\br)}$ may be called a partial functional derivative perpendicular to the integer-$N$ ``surface'' in $\rho$ space.) So when trying to solve the Euler-Lagrange equation \eqref{eq:Euler-Lagrange} we are confronted with the problem that the crucial derivative does not exist. In section II of \cite{Baerends2020Janak} a detailed discussion of the application of the Lagrange multiplier technique in such a case is given. In short, when one wants to apply the Euler-Lagrange variation method while $E_v[\rho]$ is not defined for densities outside the integer-$N$ ones, the solution is to \textit{define} $E_v$ for such densities. That can be done arbitrarily, with only the requirement of continuity, so that the derivative $\partial E_v[\rho^N]/\partial N$ at the $N$-electron $\rho^N$ exists.  But the magnitude of  $\partial E_v[\rho^N]/\partial N$ is then arbitrary. With a defined continuous $E_v[\rho]$ in the neighborhood of the feasible domain of $N$-electron $\rho$'s, the Euler-Lagrange equation can in principle be solved. The solution at a proper $N$-electron density, for which $E_v[\rho^N]$ exists according to Hohenberg-Kohn, is then obtained. The ``force of constraint'' to keep $\rho$ at $N$ electrons is then the derivative ($\partial E_v/\partial N)_{\sigma(\br)}$. This derivative follows from the chosen, essentially arbitrary, continuation of $E_v[\rho]$ in the noninteger $N$ domain at constant shape function $\sigma(\br)$. This arbitrariness does not affect the solution at the optimum $N$-electron $\rho$, $E_v[\rho^N_0]$. \\
\\
The fact that $\mu=(\partial E_v/\partial N)_{\sigma(\br)}$ is an arbitrary constant is not in any way problematic. However, Parr et al. \cite{ParrDonnelly1978} have stated that this is ``the chemical potential'' (of the electrons in a molecule). They stipulate, without further proof or derivation, that this is a physical quantity, and is characteristic of the molecule. That conflicts with the arbitrariness we noted above. One can also see that it contradicts the gauge invariance property of the external potential, which is carefully taken into account in the HK theory. Breaking $E_v$ up into the HK functional $F[\rho]$ and the external potential dependent part $\int v(\br)\rho(\br)d\br$, one finds for $v(\br)$ from \eqref{eq:Euler-Lagrange} the well-known expression
\begin{equation}
v(\br)=-\frac{\delta F[\rho_0^N]}{\delta\rho(\br)}+\mu
\end{equation}
The constant $\mu$ is always stated to reflect the gauge freedom of the local potential $v(\br)$.  That fits in perfectly with the arbitrariness of $\mu$ noted above.
However, stating that $\mu$ is a fixed constant that is characteristic for the system (``the chemical potential'') contradicts the gauge freedom.  Does the potential have to go to a given, physical, constant $\mu$ at infinity?  We know that is not the case.\\
\\
Another problem with the notion of $\mu$ being a characteristic physical quantity of a molecule (or atom) with the meaning of the chemical potential of the electrons was brought forward by Perdew et al.\ in Ref.\ \cite{PerdewParrLevyBalduz1982} (PPLB). These authors describe the following paradox or anomaly. Suppose the flow of electron density would be governed by such a chemical potential, and take two different neutral atoms at noninteracting distance, with different chemical potentials. Then a small density transfer of magnitude $\delta N$ to the atom with lowest $\mu$ would lower the energy. This will continue till the chemical potentials (which will change upon density change) will equalize. So the energy will minimize at net negative charge (possibly even noninteger) on the atom with initially lowest chemical potential and net positive charge on the other atom. This is in contradiction with physical reality where the ground state for each pair of noninteracting atoms has neutral atoms, since no electron affinity $A$ is larger than an ionization energy $I$. 
It is also a quantum mechanical reality that the ground state wave function for two noninteracting atoms would be  a product of two atomic ground states with \textit{integer} numbers of electrons. \\
The anomalous result signalled by PPLB arises from an assumption which is maybe not inherent in the concept of a chemical potential for the electrons, but is almost automatically linked with it: that the electron distribution can be considered as an electron ``fluid'', in fact consisting of very many ``particles'' each having a tiny fraction of an electron charge, whose behavior is analogous to the behavior of the very many particles in thermodynamic systems: the flow is towards a region (or a phase) with lowest chemical potential. But electrons are not like that, they cannot fracture into a myriad of smaller particles but can only jump as a complete electron. The anomaly should lead to the conclusion that this conceptual framework does not correspond to the reality. 

\begin{figure}
\includegraphics[width=7cm]{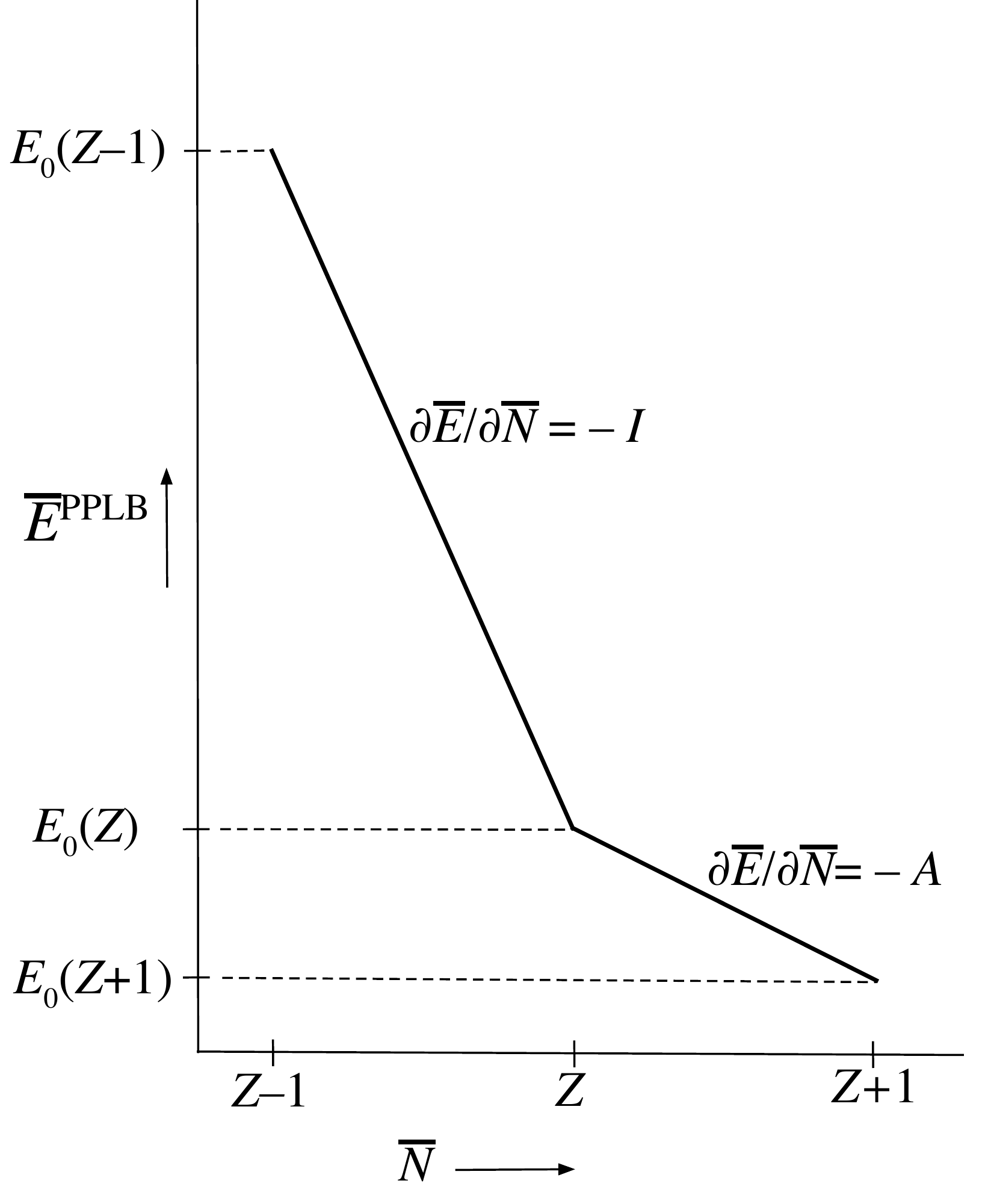}
\caption{The straight-line energy behavior as a function of noninteger electron number according to the definition of the energy on the noninteger $N$ domains by the ensemble Ansatz of Ref.\ \cite{PerdewParrLevyBalduz1982} (PPLB).  }
\label{fig:straightlines}
\end{figure}

PPLB propose a different solution of the anomaly along the following lines.  They define the energy for noninteger $N$, $E_v[\rho_{N+\omega}]$,  by making a specific choice for $\rho_{N+\omega}$ and $E_v[\rho_{N+\omega}]$, namely a linear interpolation between the (physical) ground state densities and energies of the $N$-electron system and the $(N+1)$-electron (viz. the $(N-1)$-electron) system. This is done by density and energy extension into the noninteger $N$ domain through a quantum mechanical density matrix (also called ensemble), for instance for $\overline{N}$ between $N$ and $(N+1)$ (distinguishing the noninteger $N$ by an overline), 
\begin{align}\label{eq:N+omega}
&\hat{\Dcal}=(1-\omega)\DM{\Psi^N_0}{\Psi^N_0} + \omega \DM{\Psi^{N+1}_0}{\Psi^{N+1}_0} \notag \\
&\overline{N}=N+\omega \notag \\
&\overline{E}(\overline{N})=(1-\omega)E^N_0+\omega E^{N+1}_0  \notag  \\
&\rho(\overline{N})=(1-\omega)\rho_0^N+\omega \rho^{N+1}_0  
\end{align}
Note that ``ensemble'' is used here in the quantum mechanical sense of ``mixture of states'' (to be distinguished from a superposition of states). Confusion with the  statistical mechanical (Gibbsian) ensembles to be discussed in section \ref{sec:PPLB} is to be avoided. 
Eq.\ \ref{eq:N+omega} implies linear energy behavior between the integer $N$ points, as depicted in Fig.~\ref{fig:straightlines} for an atom with electron number between $Z-1$ and $Z+1$. It is the ``straight-lines'' behavior frequently referred to that leads to the famous derivative discontinuity of the energy at integer $N$: 
\begin{align}
\left(\frac{\partial \overline{E}}{\partial \overline{N}}\right)_- = -I, \quad \left(\frac{\partial \overline{E}}{\partial \overline{N}}\right)_+ = -A 
\end{align}

This solves the paradox in the sense that a small density change $\delta N$ will now have energy increase proportional to $I$ at one atom and energy lowering proportional to $A$ at the other atom. In fact, the correct situation has been restored that either the electron will go over in its entirety or not at all, depending on the magnitudes of $I$ and $A$.  However, this continuation of the density into the noninteger domain is not in keeping with the fact that the partial derivative  $(\partial E_v/\partial \overline{N})_{\sigma(\br)}$ has to be taken with constant shape function $\sigma(\br)$. It should be stressed that $\partial E_v/\partial \overline{N}$ is a partial derivative, meaning that it has to be taken while the shape of the density $\sigma(\br)=\rho(\br)/N$\ is constant \cite{Baerends2020Janak},
\begin{equation}
\frac{\partial E_v[\rho^N]}{\partial \overline{N}}=\left[ \frac{\delta E_v[\rho^N]}{\delta\rho(\br)} \right]_{\sigma(\br)}.
\end{equation}
In order to be able to (theoretically at least) apply the Euler-Lagrange method it is required that one extends the definition of $E_v[\rho]$ to noninteger densities in such a way that $\partial E_v/\partial \overline{N}$ is a defined constant (even if that constant is not prescribed, so a lot of freedom). One cannot apply the Euler-Lagrange method if the derivative at integer $N$ does not exist, which is the case if left and right derivatives are different (then the force of constraint cannot be determined). The restriction to density changes resulting from an ensemble of two integer $N$ states necessarily makes the density change at an integer $N$ point discontinuous and therefore precludes solution of the Euler-Lagrange equation \eqref{eq:Euler-Lagrange}. \\
The most straightforward extension of the density into the nonphysical fractional electron domain while keeping the density shape constant  would be to choose $\rho^{N+\omega} \equiv (N+\omega)\sigma = (\overline{N}/N) \rho^N$ and to define the corresponding energy as $E_v[{\rho^{N+\omega}]=(\overline{N}/N)E_v[\rho^N}]$. The derivative with respect to $\overline{N}$ at constant $\sigma(\br)$ is simple and continuous at the integer $N$ point. Lieb \cite{Lieb1983} mentioned the possibility  $E_v[\rho^{\overline{N}}] = \overline{N}E_v[\rho^{\overline{N}}/\overline{N}]$ for the extension of the definition of $E_v[\rho]$  to the noninteger $N$ domain. However, for $\overline{N}$ integer this does not revert to the standard value $E_v[\rho^N]$. PPLB do not keep the shape $\sigma(\br)$ of the density constant in the neighborhood of the integer $N$ density, but make a break exactly at that point. That is what the derivative discontinuity reflects.\\
The  PPLB  straight-line energies for noninteger electron number could be called just a possible definition, since we have seen the energy for noninteger $N$ is not a physical quantity and can be defined in any way we like. It is nevertheless important that these straight-line energies are not determined by the physics of some real (existing) system. They do not represent ``the exact DFT energy for noninteger $N$'', a point to which we return below. It is one of the possible choices for the continuation of $E_v[\rho]$ into the unphysical domain of noninteger densities. Given the arbitrariness of this choice, one cannot expect that any physics can be derived from it, neither from the discontinuous PPLB choice of the derivative nor from any continuous choice.\\
\\
If a small electron density increase at an $N$-electron atom is required to have the shape of the $(N+1)$ ground state density $\rho^{N+1}_0$, and if we wish to describe the total $(N+\omega)$-electron density with a single set of Kohn-Sham orbitals, the highest energy Kohn-Sham orbital (the one with $\omega$ electrons) must have orbital energy $-A$. This is necessary because the asymptotic behavior of the $\rho^{N+1}_0$ density is known to be exponential as $e^{-2\sqrt{2A}\,r}$. At the same time the asymptotics is determined by the slowest decaying KS orbital density, which is governed by its orbital energy as $e^{-2\sqrt{-2\epsilon_L}\,r}$ (this orbital with occupation $\omega$ is the former LUMO, hence the subscript $L$). So we must have $\epsilon_L(N+\omega)=-A$. Now it is known that the exact Kohn-Sham orbital energy of the LUMO of the $N$-electron system, $\epsilon_L(N)$, is usually (for closed shell molecules) considerably lower than $-A$ \cite{Baerends2018JCP}, which can be understood from the physical nature of the KS potential \cite{BaerendsGritsenkovanMeer2013} (see section \ref{subsec:BandGap}). The implication of the prescription $\epsilon_L(N+\omega)=-A$ then is that the KS potential for any finite density increase $\delta N$ having shape $\rho_0^{N+1}$, however small, must shift up by a constant over the molecular region (so as not to disturb the shapes of the fully occupied orbitals making up the $\rho^N_0$ density) of magnitude $\Delta=-A-\epsilon_L(N)$. This jump raises all orbital levels so that the LUMO level (with now $\omega$ electrons) becomes $\epsilon_L(N+\omega)=-A$, see \cite{PerdewParrLevyBalduz1982} and \cite{PerdewLevy1983}. Note that the constant should not extend to infinity, since the KS potential must always go to zero asymptotically in order to give the orbital energies absolute meaning (not dependent on an arbitrary gauge choice). The radius $R$ beyond which the constant should no longer be effective \cite{PerdewLevy1983} and the potential has returned to the asymptotic $-1/r$ behavior can be estimated \cite{Leeuwen1996Topics}. \\
When one introduces this jumping behavior of the KS potential, the fundamental band gap $I-A$ is obviously restored if one takes the LUMO level after the jump has occurred (but the HOMO level before the jump): $\epsilon_L(N)+\Delta -\epsilon_H(N)=I-A$. We have noted that this jumping behavior of the KS potential  is not a physical phenomenon, it is only required if the density extension beyond the integer $N$ is prescribed to have the $\rho^{N+1}_0$ shape. It has nevertheless been considered to provide  an explanation for the band gap problem. We will return to this issue in section \ref{subsec:BandGap}.\\
\\
We have been concerned here with ground states, i.e.\ solutions of the Schr\"{o}dinger equation. The HK theorems have revealed that an alternative procedure to obtain the ground state energy would be the solution of the Euler-Lagrange equation \eqref{eq:Euler-Lagrange}, if the functional $E_v[\rho]$ would be known. But this does not change the quantum mechanical reality that the ground state (any energy eigenstate) can be fully known by solving the Schr\"{o}dinger equation. There are no other variables, like chemical potential or temperature, that could also affect the eigenstates. These are not a kind of ''hidden variables'' that also have to be known in order to fully characterize an eigenstate. \\
It would therefore appear that statistical mechanics has little relevance for an understanding of properties of the ground state, and of a mixture of ground states.   Statistical mechanics is just concerned with the distribution in a macroscopic system of particles like atoms and molecules over the known eigenstates. It makes us understand how this distribution can be described with thermodynamic quantities like temperature and chemical potential. Using the so-called energy representation, one pictures the particles in a gas of say electrons and  molecules (possibly ionized) as being in energy eigenstates most of the time (except for the instants where they change their state, e.g.\ by collisions, so that equilibrium can be achieved and maintained). These states do not themselves depend on the temperature or chemical potential. Only the distribution over the states is tied to these macroscopic variables. \\
Nevertheless, in DFT a connection of ground state solutions (energies, densities)  with thermodynamics has been pursued. The rationale  seems to be that the $T \to 0$ limit of a thermodynamic treatment should substantiate the concept of a chemical potential and the associated straight lines behavior of Fig.\ \ref{fig:straightlines}, together with the notion of a derivative continuity of the energy \cite{PerdewParrLevyBalduz1982,Perdew1985NATO,SagredoBurke2020}. We will consider these notions in detail in the next two sections. However, that will not change the point of view expounded in the present section, and does not have relevance for the consequences that are listed in section \ref{sec:Miscellaneous}.

\section{Atoms and molecules as thermodynamic systems?}\label{sec:PPLB}
There is frequent reference in the present day DFT literature to atoms as thermodynamic systems, with the electrons as particles. As generally the case in grand canonical ensembles, they are considered as open systems that can exchange particles (electrons) with a reservoir, the particles of the system (the electrons) having a chemical potential and a temperature that can be varied (dictated) by the reservoir. In that case the probability distribution over energies and particle numbers of the grand canonical (GC) ensemble is applicable.  Normally in the  GC ensemble all particle numbers are taken into account, but here the positive ions (up to the completely ionized  $Z+$ ion) are admitted, plus the neutral atom and the anion. So each ensemble member has a specific number $N_i$ of electrons, ranging from $N_0=0$ to $N_{Z+1}=Z+1$. For each particle number $N_i$ there is a series of energies $E_j(N_i), j \ge 0$, ranging from the ground state  ($j=0$ denotes the ground state) to some maximum excited state. For an atom with energy  $E_j(N_i)$ the GC probability contribution then is
\begin{align}\label{eq:fractionsGH}
p(N_i,E_j(N_i))&=\frac{\exp{[(N_i\mu-E_j(N_i))/kT]}}{\sum_{i=0}^{Z+1} \sum_j  \exp{[N_i\mu-E_j(N_i))/kT]} }  \\
Z^{GC} &= \sum_{i=0}^{Z+1} \sum_j  \exp{[N_i\mu-E_j(N_i))/kT]}. \notag
\end{align} 
The denominator (the GC partition function $Z^{GC}$) just takes care of normalization. The average number of electrons over this collection of ions is 
\begin{align}\label{eq:Naverage}
\overline{N}&=\sum_{i=0}^{Z+1}  \sum_j N_i p(N_i,E_j(N_i)) \\
   &=\sum_{i=0}^{Z+1} \sum_j N_i \frac{\exp{[(N_i\mu-E_j(N_i))/kT]}}{\sum_{i=0}^{Z+1} \sum_j  \exp{[(N_i\mu-E_j(N_i))/kT]} } \notag
\end{align}
Evidently $\overline{N}$ is in general a noninteger number. \\
In the same way the average energy $\overline{E}$ can be obtained as a function of $\mu$. $\overline{E}$ is defined at any $T$  as 
\begin{align}\label{eq:Eaverage}
\overline{E}(\mu)&=\sum_{i=0}^{Z+1} \sum_j p(N_i,E_j(N_i)) E_j(N_i)  \\
   &=\sum_{i=0}^{Z+1} \sum_j E_j(N_i) \frac{\exp{[(N_i\mu-E_j(N_i))/kT]}}{\sum_{i=0}^{Z+1} \sum_j  \exp{[(N_i\mu-E_j(N_i))/kT]} } \notag
\end{align}
This approach has been followed in the work of Gyftopoulos and Haftopoulos (GH) \cite{Gyftopoulos1968} and has been adopted in the DFT literature \cite{PerdewParrLevyBalduz1982,ParrBartolotti1983,Perdew1985NATO}. PPLB \cite{PerdewParrLevyBalduz1982} noted that when they only consider an average $\overline{N}$ between $Z-1$ and $Z+1$ and at the same time take the limit for $T \to 0$, only the three terms, corresponding to the atoms with charges +1, 0 and $-1$, will contribute.  They then take the GC-like probability distribution over the ground states of these systems as starting point,
\begin{align}\label{eq:PPLBp_i}
p_J=\frac{\exp{(\mu J -E_0(J))/kT}}{\sum_{M=Z-1}^{Z+1} \exp{(\mu M -E_0(M))/kT}}
\end{align}
where $J$ can take on the values $J=Z-1, Z, Z+1$. The argument for ground states only is that, comparing the terms in the $Z^{GC}$ belonging to a given particle number, it is clear that the excited states will have a contribution that at $T \to 0$ will be vanishingly small compared to the contribution from the ground state, so \eqref{eq:fractionsGH} is simplified to \eqref{eq:PPLBp_i}.\\
The GC probability distribution depends on the temperature $T$ and chemical potential $\mu$ of the electrons in the atom. In order to fix these quantities the customary device of contact with a usually unspecified ``reservoir'' is invoked that would be able to endow the electrons with these properties, and can modulate them at will, apparently without causing any essential disturbance of the (properties of the) atom. The latter requirement is important if one wants to deduce any free-atom property from this device. \\
\\
However, the few electrons in an atom do not constitute a thermodynamic system in the usual meaning of the term. If chemical potential and temperature are not properties of the system, the device of contact with a reservoir in order to bring these properties to desired values cannot be invoked. Since the notion of atoms/molecules as thermodynamic systems of electrons seems to be widespread in the DFT community, we feel it is important to dispel it. We will use arguments from elementary statistical mechanics, and refer to  Appendix \ref{app:statistics} for a brief exposition of the statistical mechanical underpinnings of thermodynamics. More detail can be found in the many excellent textbooks on the subject \cite{Tolman1938,Rushbrooke1949,Hill1956,Hill1960,Landsberg1961,Reif1965,LandauLifshitz1980,Pathria2011}.  \\
\\
The derivation of the equation for the probability distribution over the members of an ensemble, like Eq.\ \eqref{eq:fractionsGH}, proceeds in statistical mechanics in two steps (see Appendix \ref{app:statistics}). \\
In the first step it is crucial that proper statistics can be done, which requires large numbers, both of particles in the system and in the case of the grand canonical ensemble also of a very large number of systems in the ensemble. These  ensemble members represent ``microstates'' of the target thermodynamic system, which as a macroscopic system will  traverse in the course of time very many microstates compatible with the few thermodynamic variables defining its state, such as temperature $T$, volume $V$ and number of particles $N$ (or chemical potential $\mu$ in the GC case). The large numbers give rise to statistics when, to mimick the time behavior of the real system, a distribution of these microstates  over a huge ensemble of ``macroscopically identical'' systems would be constructed. This is the statistical mechanical device which, assuming equal \textit{a priori} probabilities for the micostates, leads to the distribution \eqref{eq:grandcanonicalprobability}. As always in statistical mechanics, the constraints on total number and total energy are introduced with the Lagrange multipliers $\alpha$ and $\beta$ that feature in Eq.\  \eqref{eq:grandcanonicalprobability}.  However, such microstates of ``macroscopically identical'' systems, that are traversed in the course of time,  do not exist in the case of an atom as ``thermodynamic system''. The ``laws of large numbers'' that are the basis of statistical mechanics require an enormously large number of particles $N$ in the thermodynamic system that is the target of the statistical mechanical derivations, and for the grand canonical ensemble an enormously large number $\mathcal{N}$ of systems in the Gibbsian ensemble. But here we have very few systems, with very low particle numbers. These particle numbers are very different from the average, but an overwhelmingly large number of the members of the ensemble should have particle numbers very close to $\overline{N}$. So the derivation of the probability distribution \eqref{eq:grandcanonicalprobability} cannot be carried through for an atom.\\
\\
 In the second step the Lagrange multipliers $\alpha$ and $\beta$ in \eqref{eq:grandcanonicalprobability} should be shown to have the usual physical meanings  in terms of the chemical potential $\mu$ and temperature $T$ of the thermodynamic system that the grand canonical ensemble is to represent. That allows to obtain the probability distribution in terms of $\mu$ and $T$, as in  Eq.\ \eqref{eq:gcprobability} (cf. \eqref{eq:fractionsGH} - \eqref{eq:PPLBp_i}). To make this identification, the First Law of Thermodynamics (cf. \eqref{eq:FirstLaw}) is invoked, see e.g.\ Pathria and Beal \cite{Pathria2011}  Ch. 4.3 or Hill \cite{Hill1956}, Ch. 3. So the system \textit{must be} a thermodynamic system for which the First Law (including its ingredients such entropy, temperature, chemical potential) are applicable. However, the few electrons in the target system, a free atom, do not constitute such a thermodynamic system to which the First Law can be applied. In particular, the properties of temperature and chemical potential (of the electrons) do not exist.\\
 \\
As for the chemical potential, we have noted in section \ref{sec:Euler-Lagrange} the problems that exist with the definition $\mu = \partial E/\partial N$ on account of the absence of a definition of the energy for infinitesimal increase of the particle number. The arbitrariness of the Lagrange multiplier $\mu$ in the Euler-Lagrange equation \eqref{eq:Euler-Lagrange} shows that the HK theorems do not provide a basis for the definition of a chemical potential for the electrons in an atom. More generally, there is a well-known small-number problem with defining the chemical potential if the particle number is not extremely large. We recall that the chemical potential is defined in thermodynamics as the partial derivatives in Eq.\ \eqref{eq:chemicalpotential}. Using $F=E-TS$ and $G=E+pV-TS$ one also has
\begin{align}\label{eq:dF/dN.dG/dN}
 \mu &=\left(\frac{\partial E}{\partial N}\right)_{T,S}= \left(\frac{\partial F}{\partial N}\right)_{T,V} = \left(\frac{\partial G}{\partial N}\right)_{T,p} 
\end{align}
It is clear that with a small number of particles one cannot expect the addition of a particle or the removal of a particle (with the given constraints) to have the same effect (apart from sign). The particle number has to be so large that there is virtually no difference between these two energies; for the derivatives of Eqns \eqref{eq:chemicalpotential} and \eqref{eq:dF/dN.dG/dN} to be defined the left derivative (using $\Delta N=-1$) and the right derivative (using $\Delta N=+1$) should be (practically) the same. For particle numbers in the order of Avogadro's number the change by 1 particle is in suitable systems (not in all systems!) equal for addition and removal to any desired practical accuracy (the mathematical derivative can then be  closely approximated by a finite change $\Delta N=+ 1$ or $-1$). That is not true with a small number of particles. Then one typically still is in the regime where consecutive additions of a particle do not yet involve the constant energy changes that will be reached at very large particle numbers. This is definitely the case for atoms with their wide disparity between ionization energy and electron affinity. Hill in his treatise on the thermodynamics of small systems \cite{Hill1964}  recognizes the impossibility of a single valid definition of the chemical potential at small particle numbers.  \\
Let us also note that Eqns \eqref{eq:fractionsGH} and ~\eqref{eq:PPLBp_i} not only require that $\mu$ is defined, but also that it is equal for the electrons in the neutral atom and in the positive and the negative ions.  The GC ensemble is a $(\mu, T, V)$ ensemble, i.e.\ of the pair $(\mu, N)$ it is $N$ but not $\mu$ that may vary between the members of the ensemble.  But equal $\mu$ for the electrons in all the ions is intuitively not plausible when each member of the ensemble is just an ion in an energy eigenstate. Indeed, in the DFT literature a frequent suggestion has been  that $\mu$ for an atom would be between the ionization energy $I$ and electron affinity $A$. GH and PPLB \cite{PerdewParrLevyBalduz1982,Gyftopoulos1968} derive it to be  $\mu=-(1/2)(I+A)$ when $\overline{N}=Z$ (i.e.\ in the neutral atom), which is equal to the Mulliken electronegativity, see also below. But the average of ionization energy and electron affinity is certainly very different for the neutral atom and the positive and negative ions.  Both formally and intuitively, the notion of a chemical potential of the electrons in a single free atom/ion which would be independent of its charge and common to all of its energy eigenstates is not plausible.  \\
\\
Turning then to the temperature,  it is clear one cannot measure the temperature of the electrons in an atom, and neither can one establish equilibrium by establishing equality of temperature in parts of the system. It is an elementary law of thermodynamics that one should be able to measure the temperature of the experimental system. The importance of the existence and measurability of temperature as the foremost characteristic of a thermodynamic system has eventually led to its codification into the Zeroth Law of Thermodynamics, cf.\ Reif \cite{Reif1965}. Temperature is not an ensemble property, it should be a measurable physical property of the single system for which a representative ensemble is invoked to make deductions about its equilibrium properties. For the concept and measurement of temperature it is relevant that the thermodynamic limit for the number of particles in one system can be reached, lest the temperature remains undefined at the required level of precision. In the present case the particles are the electrons of the atom. For them the thermodynamic limit of ca.\ $10^{23}$ never comes into play. We emphasize that we are not dealing with the prototypical statistical mechanical  case of a gas where the particles are atoms, in which case we have very many particles in a macroscopic volume $V$ and the  temperature $T$ is related to the occupation of the translational energy states (the kinetic energy). At higher $T$ the occupation of electronically excited states starts to play a role. The present case is very different. We would need a common temperature that can be deduced for the electrons in  energy eigenstates of an atom and its ions.  We conclude that temperature is not a property of the few electrons in an atom, let alone that it could be (made) equal in all the ions in their stationary states, as implied by the probability distributions \eqref{eq:fractionsGH} and \eqref{eq:PPLBp_i}. \\
\\
So the electrons in an atom do not constitute a thermodynamic system. Then equations  \eqref{eq:fractionsGH} - \eqref{eq:PPLBp_i}  cannot be derived as the probability distribution over the members of a GC ensemble of thermodynamic systems. Now in spite of the fact that an atom is evidently not a  thermodynamic system, it has been stated that nevertheless the electrons can derive thermodynamic properties like $\mu$ and $T$ from contact with a reservoir. Then obviously the reservoir is assigned a different role from the usual one.  In statistical mechanics a genuine macroscopic  thermodynamic system which does have properties like $\mu$ and $T$ is sometimes (thought to be) brought into contact with another huge thermodynamic system (the reservoir) which can exchange only heat with it (to establish the temperature) or both heat and particles (to establish both temperature and  chemical potential). In that way $\mu$ and $T$ of the thermodynamic system can be brought to desired values. What happens if the system is such that $\mu$ and $T$ do not exist for the isolated system? Can these properties be conferred to it in some magical way by the contact (what kind of contact?) with the reservoir? That is not the case. There are special cases where some properties can be derived from a partition function of a small subsystem (not itself a thermodynamic system)  of a large system \cite{Brielscomment,Hill1960}. This may happen if a large thermodynamic system contains many small subsystems, for instance adsorption sites on a surface in equilibrium with a gas of adsorbing particles (see e.g. Hill \cite{Hill1960}, Ch. 7). Independence of the subsystems then may lead to simplification. We discuss some examples of this special case in Appendix \ref{app:reservoir}. The conclusion remains that the electrons in an atom do not constitute a thermodynamic system and as such do not afford the definition of thermodynamic quantities like temperature or chemical potential for such a system.   \\ 
So we reject the applicability of the GC  Eqns \eqref{eq:fractionsGH} - \eqref{eq:PPLBp_i} to atoms. We will nevertheless consider the question whether these equations, if accepted, would justify the picture of Fig.\ \ref{fig:straightlines} and the related assumption that the chemical potential of the electrons exists and is $\partial \overline{E}/\partial \overline{N}$. In the next section (\ref{sec:steps}) we will investigate this in detail and will conclude negatively.  Readers who are convinced at this point that atoms and molecules are not \textit{bona fide} thermodynamic systems, and therefore Eqns  \eqref{eq:fractionsGH} - \eqref{eq:PPLBp_i} cannot provide a justification of the existence of a chemical potential for the electrons in an atom, as an atomic property, can skip these details and move to section \ref{sec:Miscellaneous} where a number of concepts in DFT are discussed in the light of the findings of section \ref{sec:Euler-Lagrange}.

\section{Step behavior of $\overline{N}$ and $\overline{E}$ as function of $\mu$ \newline and behavior of $\overline{E}$ as function of $\overline{N}$,\newline in the limit $T \to 0$ } \label{sec:steps}

\begin{figure}
\includegraphics[width=7cm]{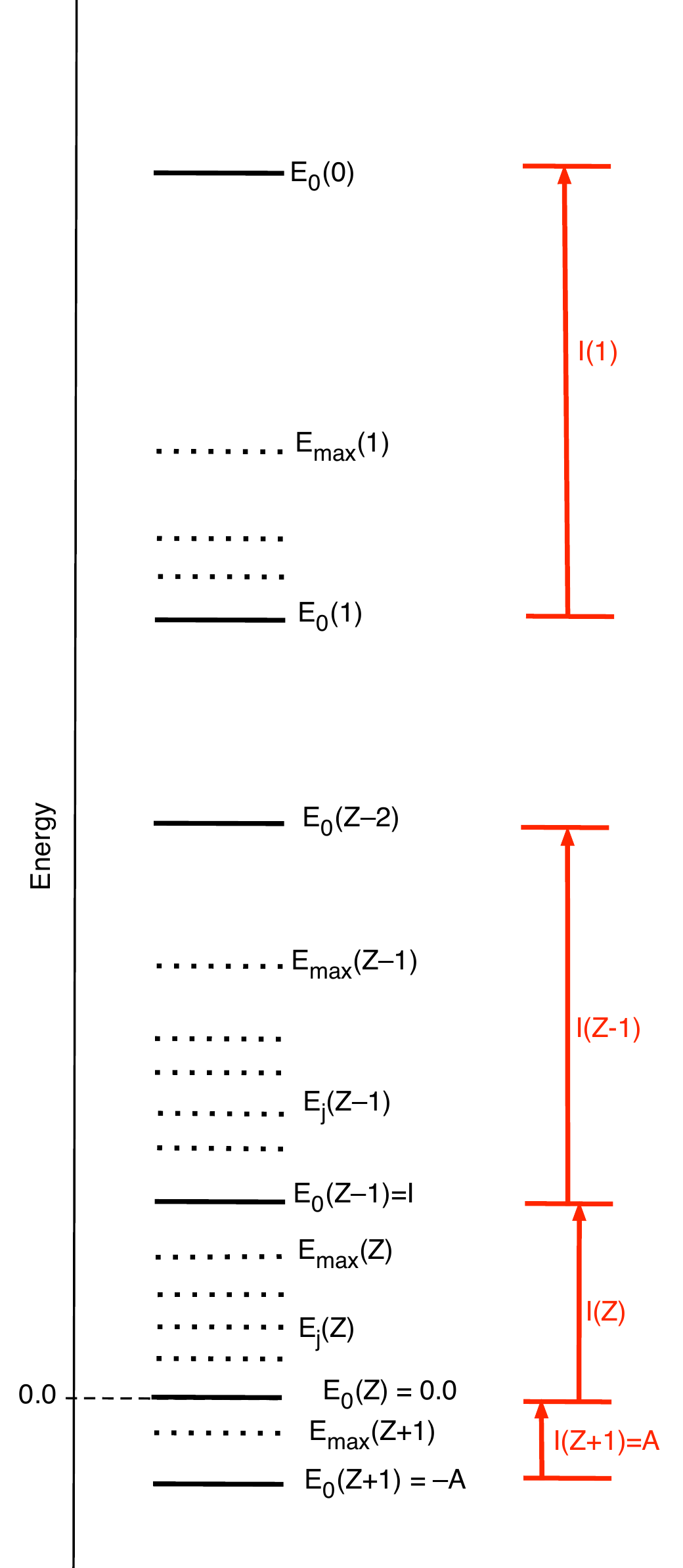}
\caption{Schematic picture of the energies of the neutral atom (energy $E_0(Z)=0.0$, i.e.\ put at energy zero), the anion at $E_0(Z+1)=-A$, the +1 ion  at $E_0(Z-1)=I$, etc. till the fully ionized system at $E_0(0)$. A number of excited states $E_j(Z^\prime),  (+1 \le Z^\prime \le 0)$, may be included, up till some maximum excited state with energy $E_{max}(Z^\prime)$. The ionization energy for a $Z^\prime$-electron ion is labeled $I(Z^\prime)$ and it is assumed, as is generally the case, that the ionization energy increases with increasing positive charge on the ion, $I(Z^\prime-1) \gg I(Z^\prime)$. }
\label{fig:EnergiesIons}
\end{figure}

\begin{figure}
\includegraphics[width=8cm]{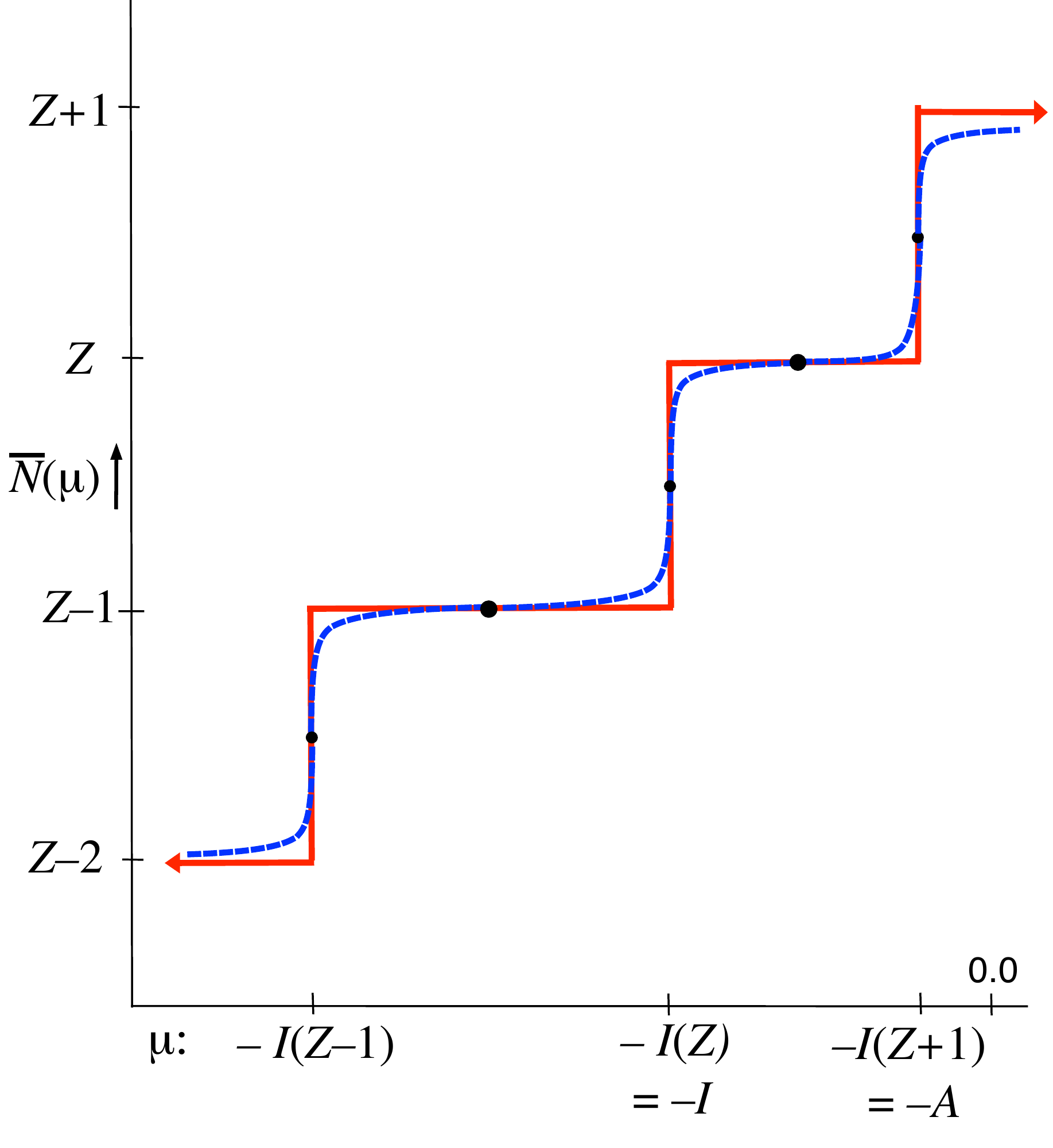}
\caption{Steps in the average electron number $\overline{N}$ as a function of the parameter $\mu$ in the limit $T \to 0$.  For $T$ small the steps are soft, see the blue dashed line. Heavy black dots: values of $\overline{N}(\mu)$ independent of $T$; small black dots: values of $\overline{N}(\mu)$ obtained in the limit $T=0$ only.}
\label{fig:NSteps}
\end{figure}

We now leave aside the question if  $\mu$ and $T$ of atomic electrons are fictitious but just want to investigate if the relations \eqref{eq:fractionsGH} - \eqref{eq:PPLBp_i} do substantiate, as proposed, the behavior of the average energy $\overline{E}$ and average particle number $\overline{N}$ as a function of the chemical potential, and in particular the behavior of $\overline{E}$ as a function of noninteger particle number $\overline{N}$ as depicted in Fig.\ \ref{fig:straightlines}.\\ 
The underlying  ``ensemble'' for Eqns \eqref{eq:fractionsGH} - \eqref{eq:PPLBp_i} is not a proper GC ensemble with very many members that represents at a given moment the time behavior of the thermodynamic system it represents.  We can consider it as a man made collection of ions  with only very few members.  $\mu$ and $T$ in Eqns \eqref{eq:fractionsGH} and \eqref{eq:PPLBp_i} can be considered as just parameters with which one can regulate the fractions $p(N_i,E_j(N_i))$ of the various ions in this collection of (a few) ions. The averages $\overline{N}$ and $\overline{E}$ are then functions of the parameters $\mu$ and $T$. For particular extreme choices of the parameters, like $T \to 0$ or $\mu = \infty$, particular results for the $p_i$ and the averages result. These are reviewed in this section. \\
As first example of the parameter tuning Ref.\ \cite{Gyftopoulos1968} mentions the extreme choice $\mu=-\infty$ for which all terms with $N_i > 0$ lead to negligible contributions for any finite $T$. Then only $N_i=0$ survives and $\overline{N}=0$, corresponding to the fully ionized atom. The collection of ions then has effectively one member. Another extreme parameter choice would be $\mu=+\infty$, which makes the term $i$ with maximum number of particles $N_i=Z+1$ overwhelmingly larger than any other term, hence $\overline{N}=Z+1$, corresponding to the anion. These cases where the ``ensemble'' has only one member constitute extreme deviations from the statistical mechanical ensembles. By  varying $\mu$ one can vary $\overline{N}$ between these extremes of 0 and $(Z+1)$.\\
The parameter $T$ in the expression \eqref{eq:fractionsGH} for the fractions of ions in the collection occurs in the denominator of the arguments of  exponential functions. This causes the $T \to 0$ limit to have the extreme effect of blowing up the arguments. This causes just one term in the sum \eqref{eq:Naverage} for $\overline{N}$  to be overwhelmingly larger than all other ones, namely the one with the largest (positive or least negative) argument. $\overline{N}$ will be equal to the $N_i$ of the dominating term. So in the limit $T \to 0$ $\overline{N}$ will exhibit step behavior as a function of $\mu$, making a jump when $\mu$ crosses a boundary to an interval with another dominating term. Figure \ref{fig:EnergiesIons} gives a picture of the energies of the various ions, both the ground states $E_0(Z^\prime)$, as well as the possibly included excited states $E_j(Z^\prime)$, with the maximum included excited state energy for $Z^\prime$ denoted $E_{max}(Z^\prime)$. (Here and in the sequel we denote with $Z'$ any of the integer values from $Z+1$ to 0.) The largest exponent of the terms for $Z^\prime$ particles is the one involving the ground state energy, since always $Z^\prime\mu-E_0(Z^\prime))> Z^\prime\mu-E_{j>0}(Z^\prime)$. The threshold where $\overline{N}$ switches from $Z^\prime-1$ to $Z^\prime$ is for the $\mu$ making this largest term for $Z^\prime$  larger than the largest of the $(Z^\prime-1)$ terms:
\begin{align}
Z^\prime\mu-E_0(Z^\prime) & > (Z^\prime - 1)\mu -E_0(Z^\prime-1)  \notag \\
\mu & > -(E_0(Z^\prime-1)-E_0(Z^\prime))=-I(Z^\prime)  
\end{align}
 At this point the largest $(Z'+1)$ term is still smaller than the largest $Z'$ term, the $(Z'+1)$ terms only taking precedence when
\begin{align}
\mu > -I(Z^\prime+1)
\end{align}
and we assume throughout that, as is invariably the case, the ionization energies increase with increasing charge on the system, $I(Z^\prime) > I(Z^\prime+1)$. (Note that we are in a regime with negative $\mu$.) So each time $\mu$ passes a $-I(Z')$ boundary value we expect a jump up of $\overline{N}$ from $(Z'-1)$ to $Z'$. This step behavior of $\overline{N}(\mu)$ can be read from Fig.\ \ref{fig:NSteps}. It is to be noted that the presence of excited states does not affect the steps, if the ground states are present in the collection of ions.\\
\\ 

At fixed $T$ the average $\overline{N}$ only depends on the $\mu$ parameter. The blue dashed curve in Fig.\ \ref{fig:NSteps}  depicts the steps in the function $\overline{N}(\mu)$ at small $T$, which are well known \cite{Perdew1985NATO,SagredoBurke2020}. In the limit $T \to 0$ the straight-line steps are approached. It is easy to deduce from Eq.\ \eqref{eq:Naverage} that the width of the jumps at the $\mu=-I(Z')$ ordinates of Fig.\  \ref{fig:NSteps}, is of the order $kT$: for $\mu=-I(Z') + \delta/(kT)$ $\overline{N}$ goes to $Z'$ at $\delta \gg kT$ and to $Z'-1$ at $\delta \ll -kT$. At  exactly $\mu=-I(Z')$ the exponential terms for $(Z'-1)$ and $Z'$ contribute equally and one has $\overline{N}=Z'-(1/2)$ (see small black dots in Fig.\ \ref{fig:NSteps}). This equality is not mathematically exact, since other exponential tems in \eqref{eq:Naverage} will still contribute, even if by exceedingly small amounts. On the intervals $-I(Z')< \mu < -I(Z'+1)$ $\overline{N}$ becomes constant in the limit $T \to 0$. For the midpoints $\mu=-$[$I(Z')+I(Z'+1)$]$/2$ one can analytically derive that they are practically independent of $T$, as is also intuitively obvious (see heavy black dots in Fig.\ \ref{fig:NSteps}). For the point $\mu=-(I+A)/2$ one has $\overline{N}=Z$. This has led to the suggestion that the chemical potential of the electrons in the $Z$-electron (neutral) atom would be $-(I+A)/2$, as noted above.  \\ 
\\

\begin{figure}
\includegraphics[width=8cm]{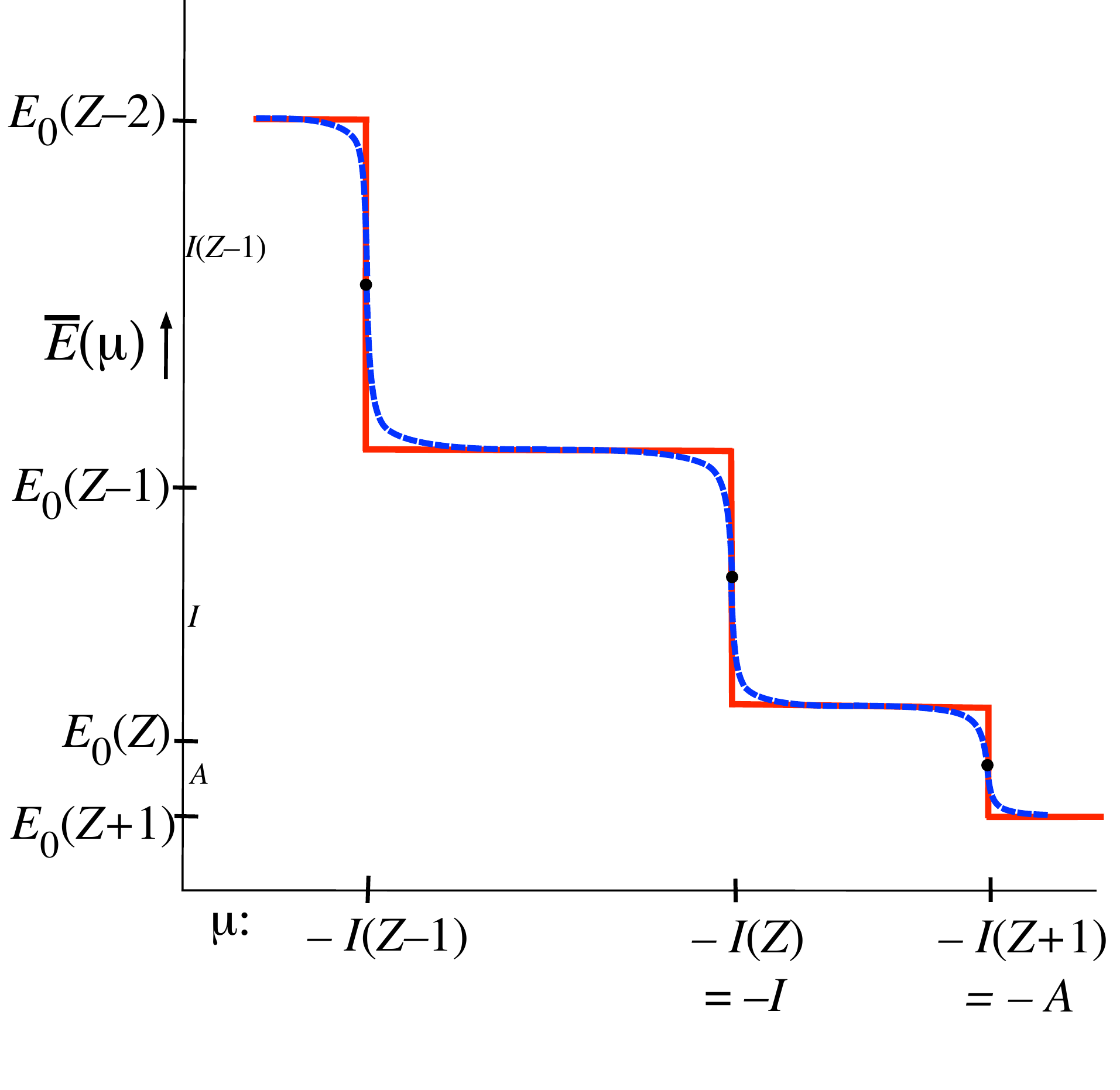}
\caption{Step behavior of $\overline{E}(\overline{N})$ as a function of the parameter $\mu$ for small $T$ (blue dashes) and in the limit $T \to 0$ (red lines). For small $T$ the steps are soft. Heavy black dots: values of $\overline{E}(\mu)$ independent of $T$; small black dots: values of $\overline{E}(\mu)$ obtained in the limit $T \to 0$ only. }
\label{fig:ESteps}
\end{figure}

The function $\overline{E}(\mu)$ \eqref{eq:Eaverage} has similar staircase behavior as the function $\overline{N}(\mu)$, with steps at the $\mu=-I(Z')$ values that cause a switch of the dominant exponential term in \eqref{eq:Eaverage}, see Fig.\ \ref{fig:ESteps}. There are some differences of detail. For instance at the midpoints $\mu=-(1/2)(I(Z')+I(Z'+1))$ the value $E_0(Z')$ is not practically independent of $T$ (no heavy black dots) but the limiting $E_0(Z')$ value is approached from above when $T \to 0$. In the limit $T \to 0$ the steps become sharp, similar to the situation for $\overline{N}(\mu)$. \\

\begin{figure}
\includegraphics[width=8cm]{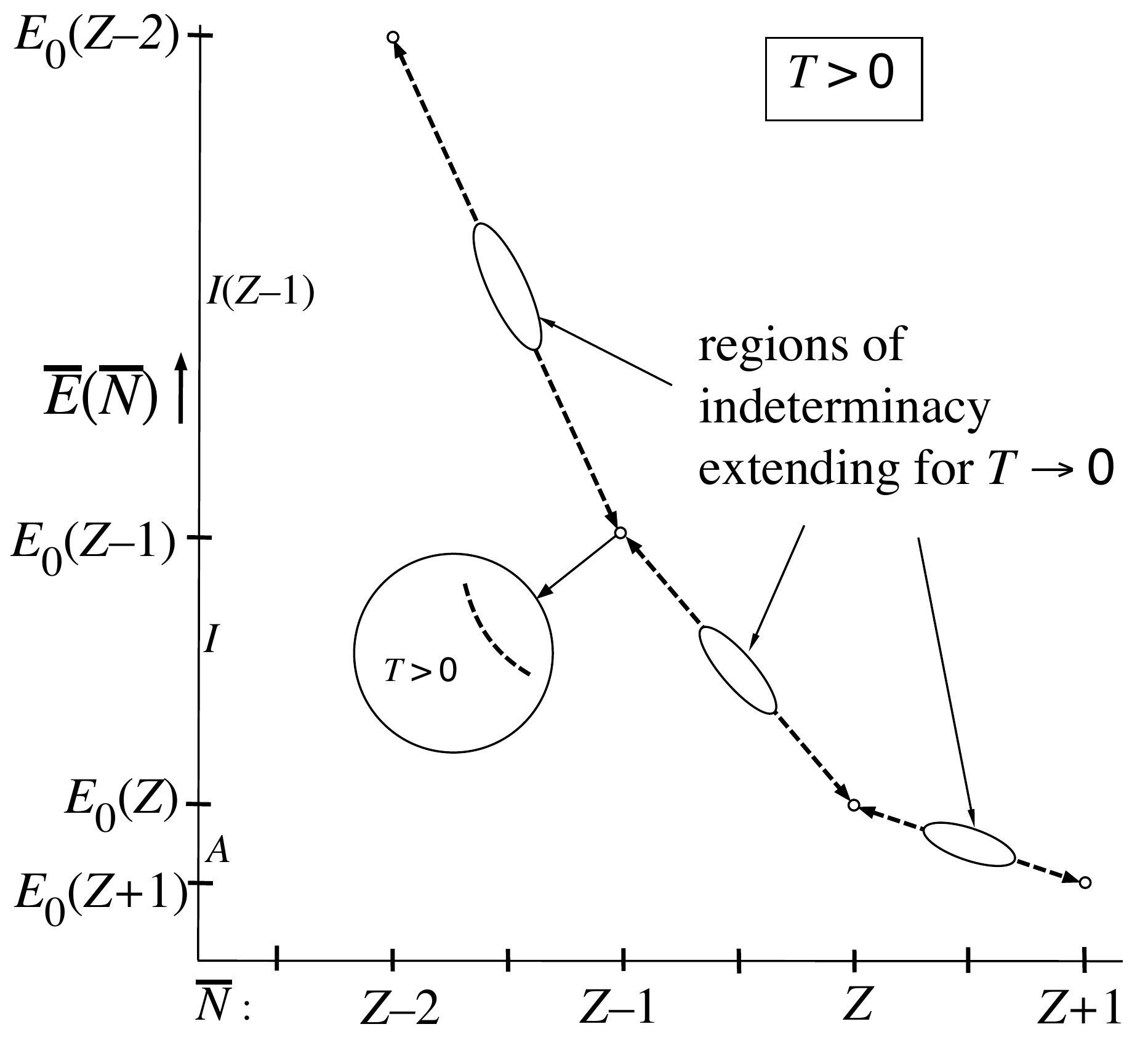}
\caption{The function $\overline{E}(\overline{N})$ for small finite $T$. The straight dashed lines have a region around $Z'-1/2$ ($Z'$ any of the integer $\overline{N}$ values) where they are not defined when $T$ becomes very small, see text. These regions extend when $T \to 0$, the lines disappearing in the limit of $T=0$, leaving only $E_0(Z')$ points at the integers $Z'$. In the regions of the small circles, around the integer $Z'$ values, the lines are, for small $T$, continuous curves with the slope changing from $-I(Z')$ to $-I(Z'+1)$ (see inset). They collapse to the point $\overline{E}(Z')=E_0(Z')$ at $T=0$, which is indicated with arrows. }
\label{fig:EStraightlines}
\end{figure}

From the figures and data for $\overline{N}(\mu)$ and $\overline{E}(\mu)$ one may deduce the behavior of $\overline{E}$ as a function of $\overline{N}$, $\overline{E}(\overline{N})$. If $\overline{N}(\mu)$ is invertible so that $\mu(\overline{N})$ is defined, $\overline{E}(\overline{N})$ can be obtained. As long as $T$ is still finite the situation is as sketched with the blue curves, so $\mu(\overline{N})$ is defined. It can be derived that  $\overline{E}(\overline{N})$ becomes a straight line with slope $-I(Z')$ on the interval $Z' -1 < \overline{N} < Z'$ ($\mu$ in the neighborhood of the $-I(Z')$ point). Considering next the $-I(Z') < \mu < -I(Z'+1)$ interval, it is clear one has $\overline{N} \approx Z' $ and at the same time $\overline{E} \approx E_0(Z')$. The limit $T \to 0$, where the steps become sharp, has to be carefully considered. The whole $-I(Z') < \mu < -I(Z'+1)$ interval leads to the single point $\overline{N}=Z'$, and the single value $\overline{E}(Z')=E_0(Z')$. At the point $\mu=-I(Z')$ $\overline{N}$ is in the range $(Z'-1,Z')$, but undetermined, and also $\overline{E}$ becomes undetermined on the range $(E_0(Z'),E_0(Z'-1))$. This is depicted in Fig. \ref{fig:EStraightlines} by the dashed lines on the $((Z'-1),Z')$ interval. It should be noted that the lines disappear for $T \to 0$. Machine calculations cannot capture this behavior because of the limited precision, but they exhibit breakdown first  in the neighborhood of the midpoints $\overline{N}=Z'-1/2$, which is understandable in view of the $\overline{N}(\mu)$ and $\overline{E}(\mu)$ curves being steepest there (both derivatives, $\partial \overline{E}/\partial \mu$ and $\partial\overline{N}/\partial \mu$ are going to $\infty$ there and their ratio will  be undetermined.). In Fig.\ \ref{fig:EStraightlines}  we indicate the growing regions of indeterminacy of $\overline{E}(\overline{N})$. At small but finite $T$ the function $\overline{E}(\overline{N})$ persists at $\overline{N} \approx Z'$, going smoothly from $E_0(Z')+\delta$ to $E_0(Z')-\delta$ and the derivative $\partial \overline{E}/\partial \overline{N}$ smoothly changing from $-I(Z')$ to $-I(Z'+1)$, see inset of Fig.\ \ref{fig:EStraightlines}. However, in the limit $T \to 0$ the whole  $\overline{E}(\overline{N})$ ``curve'' collapses to just the points $(Z',E_0(Z'))$, which is indicated with arrows on the dashed lines. It is not surprising that Eq.\ \eqref{eq:fractionsGH} leads to this collapse at $T \to 0$: In that limit the dominance of just one term in \eqref{eq:fractionsGH} becomes absolute, so (depending on the $\mu$ interval) only one ion ($\overline{N}=Z'$, $\overline{E}=E_0(Z')$) has fraction 1.0. The jumps from one dominating term to the next become discontinuous.\\

For finite $T$ the $\overline{E}(\overline{N})$ picture of Fig.\ \ref{fig:EStraightlines} has some resemblance to Fig.\ \ref{fig:straightlines}. However, the latter is for $T=0$ (or rather is temperature-less) while Fig.\ \ref{fig:EStraightlines} becomes very dissimilar to Fig.\ \ref{fig:straightlines} in the limit $T \to 0$, reducing to just single points. The straight lines are then nonexistent. \\
\\
The ``ensemble'' on which Eqns \eqref{eq:fractionsGH} - \eqref{eq:Eaverage} are based \cite{PerdewParrLevyBalduz1982,Perdew1985NATO,SagredoBurke2020} is not a proper Gibbsian ensemble. Such an ensemble has very many members, which all have the same $\mu$ and almost all a particle number $N$ that is close to the ensemble average $\overline{N}$. It represents  the time behavior of a target thermodynamic system. Here there is not a \textit{bona fide} target thermodynamic system, with very many particles that allow definition of the chemical potential $\mu$ and temperature $T$.  The ``ensemble'' here has few members with particle numbers rather different from the average $\overline{N}$ and with so few particles that $T$ and $\mu$ are not defined. So the target thermodynamic system that the Gibbsian ensemble aims to represent is actually nonexistent in this case.  The application of thermodynamic relations therefore has to be viewed with reservation. This has nevertheless been undertaken  \cite{Perdew1985NATO,SagredoBurke2020} and is briefly discussed here. The denominator of Eq.\ \eqref{eq:fractionsGH} is treated as the grand canonical partition function $Z^{GC}$, although it does not qualify as such because Eq.\ \eqref{eq:fractionsGH} cannot be derived for an atom (the Lagrange multipliers $\alpha$ and $\beta$ of Appendix \ref{app:statistics} cannot be written in terms of $\mu$ and $T$ since these do not exist for the electrons in an atom). The usual thermodynamic relations for the internal energy $E$ and the Helmholtz free energy $F$ have nevertheless been applied to such small electronic systems,
\begin{align}\label{eq:EFG}
&E=TS-pV+\mu N  \notag \\
&F=E-TS=-pV+\mu N=\mu N-kT \ln{Z^{GC}} 
\end{align}
The thermodynamic free energy applies to macroscopic thermodynamic systems, its meaning for the electrons in an atom is dubious at best. But for $T=0$ $F$ is equal to the energy $E$ since the $TS$ term will be zero because $T=0$ (and $S=0$ too, cf.\ the Third Law). So using the free energy $F$ should give the same result as $E$. The ensemble average $\overline{E}$ should be equal to the $E$ of the target thermodynamic system. But such a target thermodynamic system, with defined $E$ and $N$ (and $\mu$ and $T$) does not exist here. Using $Z^{GC}$ (see Eq.\ \eqref{eq:fractionsGH}) the statistical mechanical relation
\begin{equation}
\overline{N}(\mu,T)=kT\frac{\partial Z^{GC}}{\partial \mu}
\end{equation}
yields just the expression \eqref{eq:Naverage} for $\overline{N}$.
When we consider  $\mu$ as a parameter, as before, and use $\overline{E}(\mu)$ (Fig.\ \ref{fig:ESteps}) and $\overline{N}(\mu)$ (Fig.\ \ref{fig:NSteps}) and the inverse $\mu(\overline{N})$, the behavior $\overline{E}(\overline{N})$ as depicted in Fig.\ \ref{fig:EStraightlines} results in the limit $T \to 0$. Specifically, at very small but still finite $T$, where the inverse $\mu(\overline{N})$ can still be determined, the straight-line picture will be obtained. But at $T=0$ this breaks down: when $\mu$ passes a point like $-I(Z')$ the single dominating term in $Z^{GC}$ at $T=0$ switches from the $Z'$ term $\exp{[(\mu Z' -E_0(Z'))/kT]}$ to the $Z'+1$ term, and 
\begin{equation}
F=\mu N -kT\ln{e^{(\mu Z' -E_0(Z'))/kT}}=E_0(Z')
\end{equation}
switches to $E_0(Z'+1)$, exactly as was observed  for the behavior of $\overline{E}$.\\

As stated earlier, the present (pseudo-)thermodynamic discussion based on the ```ensemble'' Eqns \eqref{eq:fractionsGH} - \eqref{eq:PPLBp_i} cannot be expected to shed light on the (temperatureless, quantum mechanical)  Fig.\ \ref{fig:straightlines} and its implications. The present finding that at $T=0$ there is nothing but the $E_0(Z')$ energies is wholly satisfactory.  It does not support the  concept of atoms with noninteger number of electrons and an energy that is not an eigenvalue of the Hamiltonian but somewhere in between. The opinion that this is a meaningful concept appears to have settled in the DFT community and has given rise to the frequent reference to fractional electron systems, with apparently the feeling that the statistical mechanical theory of  grand canonical ensembles would condone such a concept.  \\

 \section{Miscellaneous}\label{sec:Miscellaneous}
The topics we have been discussing touch on a number of issues in DFT. In this section we review a number of those to see what conseqences follow, if any.

  \subsection{Steps of the Kohn-Sham potential}\label{subsec:potentialsteps}
 At the end of section \ref{sec:Euler-Lagrange} we concluded that the jump of the KS potential when $\overline{N}$ passes an integer value is not ``physical''. It only occurs when a discontinuous density change is postulated, as in the PPLB description of fractional electron systems. Such step behavior of the KS potential would actually preclude a stable self-consistent solution to be reached for the dissociated situation of a heterogeneous diatomic molecule (the situation of two different open-shell atoms at very large distance). This has been described in detail in section IVD of Ref.~\cite{Baerends2020Janak} and underlines that his step of the KS potential (called the derivative discontinuity step, or just the derivative discontinuity, see subsection \ref{subsec:BandGap}) must be unphysical.\\
   For completeness we mention there are also true, physical, steps in the KS potential for \textit{integer} electron systems. ``Physical'' then means: required in the potential of a noninteracting particle system in order to endow it with a density equal to the one of the interacting electron system. A very well known step is the one occurring between two atoms $A$ and $B$ (or larger fragments) with each a single valence electron so they can form a covalent bond. At large distance the single electron level at atom $B$ at the lower energy ($-I_B$) has to move up to the higher level at $-I_A$ in order to form a doubly occupied orbital with 50-50 mixing so the atoms will get the correct amount of one electronic charge density each. The KS potential therefore must form a plateau over the region of atom B of height $I_B - I_A$. This leads to a step of this height in between the atoms, as was recognized by Almbladh and von Barth  \cite{AlmbladhvBarth1985} and Perdew \cite{Perdew1985NATO}. This qualitative argument  is confirmed by an analysis of the exact KS potential. It can be shown that the so-called response part of the KS potential generates the plateau mentioned above, with exactly the height $I_B - I_A$ \cite{Gritsenko1996}. This behavior is directly related to the conditional amplitude, i.e.\ it is a direct consequence of the (strong) left-right correlation in a (weak) covalent bond. It has been studied for model systems by Maitra et al.\ \cite{Maitra2009JCTC}, also for the TDDFT case \cite{Maitra2012KSpotTDDFT,Maitra2014}, and for the case of strongly correlated systems by Giarusso et al.\ \cite{Giarusso2018}. The response part of the KS potential also has step behavior when going in an atom from one shell to the next \cite{KriegerLiIafrate1992,GritsenkoBaerendsvLeeuwen1994,GritsenkoLeeuwenLenthe1995}.

\subsection{The Kohn-Sham  band gap problem}\label{subsec:BandGap}
It is generally stated that the PPLB picture ``explains the band gap problem''. The problem is the following. In a solid the KS LUMO orbital energy of the $N$-electron system $\epsilon_L(N)$ (bottom of the conduction band)  is generally below $-A$. Since the highest occupied KS orbital $\epsilon_H(N)$ (the top of the  valence band) is (in exact KS) equal to minus the ionization energy, $\epsilon_H(N)=-I$, this means that generally the KS band gap $\epsilon_L(N)-\epsilon_H(N)$ differs from the fundamental gap $I-A$. The discrepancy may actually be large.  This is called ``the band gap problem'' because, apparently, the expectation has been that in exact KS $\epsilon_L(N)=-A$ would hold. But it does not, neither in exact Kohn-Sham nor with most DFAs, which exhibit orbital energy gaps not so different from the exact KS model \cite{GruningMariniRubio2006}. The DFAs erroneously shift \textit{all} occupied and unoccupied valence levels up by several eV \cite{GritsenkoBaerends2016,Baerends2018JCP}, although the Rydberg levels in small molecules considerably less \cite{vanMeerGritsenkoBaerends2014b}. \\
However, the PPLB picture does not \textit{explain} the band gap problem nor solves it. What has caught the attention is that in the PPLB picture the KS potential for their density $(1-\omega)\rho_0(N)+\omega\rho_0(N+1)$ has for any $\omega>0$, however small,  to jump up from the one for the $\rho_0(N)$ shape by a constant $\Delta=-A-\epsilon_L(N)$ over the molecular (or crystal) region (but not in the asymptotic limit), see end of section \ref{sec:Euler-Lagrange}. This jump is required in order to move $\epsilon_L$ up to $-A$ so that the asymptotic decay of the $\omega$ electron density in the LUMO orbital will have the proper decay (of the $\rho_0(N+1)$ charge density). Of course $-A-\epsilon_L(N)$ is just the band gap ``deficit''. The deficit of the KS orbital energy gap, or band gap, $\epsilon_L(N)-\epsilon_H(N)$, i.e.\ the difference between it and the fundamental gap $I-A$, and the jump of the potential to which it is equal, are always called derivative discontinuity (DD) in solid state physics.  (Then of course the DD (i.e.\ $\Delta$)  is a different quantity than the discontinuity in the derivative of the energy, which is $I-A$).\\ 
But there is a big if: \textit{if} $\epsilon_L(N)$ is below $-A$ the band gap problem exists and the jump of the KS potential has to occur if one wants the build up the $\rho^{N+\omega}$ density by an admixture of some $(N+1)$-electron density $\rho_0(N+1)$ to the $N$-electron density: $\rho^{N+\omega}=(1-\omega)\rho_0(N)+\omega\rho_0(N+1)$. But PPLB do not predict that $\epsilon_L(N)$ is not equal to $-A$ and do not give an estimate of the magnitude of the discrepancy and the necessary potential step. In order to understand the band gap problem one has to understand \textit{why} the KS potential leads to a LUMO level that is below $-A$. That understanding does not follow from Ref.~\cite{PerdewParrLevyBalduz1982} or Ref.~\cite{PerdewLevy1983}. It should follow from an understanding of the physics of the KS electrons, i.e.\ from the nature of the KS potential and the one-electron energies that follow from it. It is indeed perfectly understandable why the exact KS potential leads to a LUMO level that is below $-A$, see the arguments in Ref.\ \cite{BaerendsGritsenkovanMeer2013}, notably its Figs~3,4.  In short, the exact KS potential incorporates the attractive potential of the full exchange-correlation hole also for the virtual orbitals, which is not the case in the Hartree-Fock model, which does have $\epsilon_L^{HF}(N) = -A^{HF}$ (which is $\approx -A$, although in poor frozen orbital approximation). Actually, understanding the origin of the difference $-A-\epsilon_L(N)$ from the nature of the KS potential leads to a correction that can easily be calculated for solids (extended systems). It can be proven  \cite{Baerends2017} that the correction $-A-\epsilon_L(N)$  is in a macroscopic solid equal to the expectation value for the LUMO orbital (the state at the bottom of the conduction band) of the response part of the KS potential. The latter can be reasonably well approximated by the expression of Ref.\ \cite{GritsenkoLeeuwenLenthe1995} (GLLB), explaining the success of Kuisma et al.\ \cite{KuismaRantala2010} and others \cite{Thygesen2012PRB,Thygesen2015AdvEnMat,Tranblaha2018,TranBlaha2021} with this correction.\\
 PPLB have straight-line energies that have $-A$ as derivative at the $N+\omega$ side. So $A$ occurs in their picture. This does not explain why $\epsilon_L(N)<-A$, and neither does it give a strategy for the calculation of the discrepancy. One would still have to calculate or approximate $A$ in order to know the discrepancy (the DD), i.e.\ to know $-A-\epsilon_L(N)$. So one has to \textit{calculate} the fundamental gap $(I-A)$ in order to obtain the magnitude of the potential step of PPLB. There is not an independent way of establishing the derivative discontinuity DD and from there obtain the correction $-A-\epsilon_L(N)$.\\
 Such a calculation of $A$ (and $I$)  is actually quite feasible. It has been pointed out by G\"orling and coworkers \cite{Gorling2015,Gorling2016}  that it is possible to calculate the total energy differences for ionization from a periodic crystal ($I$) or addition of an electron to a crystal ($A$) from total energy calculations by series of calculations with standard band structure codes. The proposed procedure has been illustrated and applied by Tran et al. \cite{TranBlaha2019TotEn}. It has also been used to confirm \cite{PerdewYangBurkeGross2017} that for those approximations (the LDA, GGA and meta-GGA functionals) for which the Slater relation $\partial E^{appr} / \partial n_i = \epsilon_i$ holds, the orbital energy gap $\epsilon^{appr}_L(N)-\epsilon^{appr}_H(N)$ is equal to the total energy based fundamental gap $I^{appr}-A^{appr}$ (which is not the case for exact KS for which indeed Slater's relation (often called Janak's theorem) does not hold). The equality $\epsilon^{appr}_L(N)-\epsilon^{appr}_H(N) =I^{appr}-A^{appr}$ does not solve the ``band gap problem'' since $I^{appr}-A^{appr}$ and therefore $\epsilon^{appr}_L(N)-\epsilon^{appr}_H(N)$ is wrong (very different from the exact $I-A$) for the LDA and GGA functionals. The error is due to the error of these approximations for the total energy of delocalized ion states, see Ref.\ \cite{Baerends2018JCP} for detailed discussion.

 \subsection{Atoms and molecules as open systems with fluctuating electron number?}\label{subsec:fluctuation}
 In the grand canonical ensemble the particle number is not fixed for the members of the ensemble. So one may consider the fluctuation of the particle number over the ensemble. It is an important result of statistical mechanics that this fluctuation is insignificant. The same holds for the energy fluctuation, which will occur in both the grand canonical and the canonical ensemble. This very small fluctuation is generally cited to justify that thermodynamic systems can be described by any type of ensemble, the choice being dictated by considerations of (mathematical) convenience. As mentioned earlier (see also Appendix \ref{app:statistics}), one may envisage a grand canonical ensemble as a collection of a very large number of systems, each connected to a reservoir with which it can exchange particles and energy. One may also envisage a grand canonical ensemble by inserting fictitious walls in for instance a macroscopic volume of gas, which are permeable for particles and heat. The number of particles in the ``central'' partition will fluctuate over time, which is reflected in the variation of the particle number over the members of the ensemble at a given time. \\
The terminology ``open system'' and ``fluctuating particle (electron) number'' has  made its way into the density functional literature, but then not regarding thermodynamic systems, but mostly referring to the electrons in an atom. The atom is not a thermodynamic system, and the fluctuation must be of a very different type than the phenomenon treated in statistical mechanics. Usually interaction with an ``environment'' is held responsible for the fluctuation. The environment is typically just the other atoms in a molecule, or a solid surface to which the atom may be bound. We wish to stress that in the ground state of such a system (or any energy eigenstate) we are not dealing with any fluctuation phenomenon. The electron density surrounding (the nucleus of) an atom is stationary in the ground state or an excited state. This also holds when the atom is only very weakly interacting with the rest of the system, be it the remainder of the molecule from which it dissociates, or the solid surface from which it detaches. The electrons in such an atom are not like the particles of a thermodynamical system for which the phenomenon of (energy or particle)  fluctuation is well studied, cf.\ \cite{Hill1956} Ch. 3, or \cite{Pathria2011} Ch. 3.6 and 4.5. These remarks pertain to the stationary states, the eigenstates of the Hamiltonian. At elevated temperatures we need to consider a Boltzmann distribution over the states. This does not alter this statement on the lack of fluctuation in an energy eigenstate.

\subsection{The deviation-from-straight-lines error (DSLE)}\label{subsec:DLSE}
We have rejected the physical basis of the straight-lines picture of the energy for fractional electrons of Fig.\ \ref{fig:straightlines}.  Still the straight-lines energy behavior has a distinct advantage in one particular case: when a \textit{local} functional is used with this straight-line behavior in a (nearly) dissociated system of fragments with in total an uneven number of electrons, so that  fragments with a noninteger number of electrons arise. This can be seen as follows. \\
The prototypical example is H$_2^+$, but other well known examples are He$_2^+$, $[$H$_2$O-H$_2$O$]^+$ etc. (for simplicity we take identical fragments as example). The poor behavior of the LDA and GGA functionals ((semi-)local DFAs in general) in such cases was well known from the treatment of ionization from equivalent sites in a molecule  \cite{Noodleman1982PostEJB}, e.g.\ $1s$ core holes in homonuclear diatomics like He$_2$, N$_2$ or C$_2$H$_4$ and subvalence ligand levels in TM complexes like Cr(CO)$_6$.  It has for instance been highlighted in 1982 by Noodleman et al.\  \cite{Noodleman1982PostEJB} for N$_2^+$ and He$_n^+$, in 1997 by Bally and Sastry \cite{Bally1997} for H$_2^+$ and He$_2^+$, in 1999 by Sodupe et al.\ \cite{Sodupe1999} for  $[$H$_2$O-H$_2$O$]^+$ and in 2008 by Cohen, Mori-S\'anchez and Yang \cite{Cohen2008Science}  for H$_2^+$. The root cause of the problem is that the local approximation is applied in situations where non-locality is essential. For a system with a noninteger electron number on separated (noninteracting) fragments (for instance two $(1/2)$ electron charges on individual H's for long distance H$_2^+$), the local approximation causes the functional to be effectively evaluated for each fragment, i.e.\ for a noninteger electron number. But for such systems the HK functional is not even defined. Of course if the local functional would yield for each $(1/2)$ electron density just half of the required H atom energy, the total energy would still be correct. But it has been clear \cite{Noodleman1982PostEJB,Sodupe1999} that the local approximations, which all have a basic LDA exchange ingredient of $\rho(\br)^{4/3}$ in the xc energy density, do for that reason not exhibit the right scaling behavior for the correct total energy to result. If we extend the example to $n$ noninteracting fragments of $N$  electrons each, with a surplus  1 electron that will be distributed over the $n$ sites, it is clear that a local functional will have to deal with $n$ fractional electron charges of $N+1/n$ each. It has been observed \cite{YangZhangAyersPRL2000,MoriCohen2014} that a local functional yields the right energy for any $n$ if the scaling of the local energy density would be perfectly linear between $N$ and $N+1$ electrons on a fragment. This is not a proof that the behavior of the straight-lines picture of Fig.\ \ref{fig:straightlines} is correct physics. It is simply making the local approximation work in this special case, where in fact the local approximation  is not warranted, being applied to a case where nonlocal effects are vital (because the fragment systems are entangled), see Ref.~\cite{Baerends2020Janak} for further discussion. Applying an exchange-correlation functional to a noninteger electron system means that the functional is applied outside the domain of densities on which the HK functional has been defined.  \\

Let us consider a small increase $\delta N=1/n$ of the electron number on a fragment over the integer number $N$ (e.g.\ when the number $n$ of noninteracting fragments in the example above is large). A local functional will derive the energy of a fragment from the local $(N+1/n)$ electron number and the energy increase according to the straight-lines picture would be 
\begin{equation}
\delta E= \left(\frac{\partial E^{PPLB}}{\partial N}\right)_+\delta N = -A/n
\end{equation}
 which would yield the correct energy change $-A$ when summed over all fragments (the additional electron can go to any fragment, so there are $n$ degenerate wavefunctions each describing the additional electron on one site, all at energy $-A$; a linear combination with $1/n$ electron per site also has energy $-A$). It is natural \cite{YangCohenmori2012} to associate the PPLB derivative at the electron-addition side with the derivative when an infinitesimal charge is added to the LUMO,
 \begin{equation}
 \left(\frac{\partial E^{PPLB}}{\partial N}\right)_+ =\frac{\partial E}{\partial n_L}
 \end{equation}
 ($L$ stands for LUMO).
 It has been pointed out by Yang, Mori-S\'{a}nchez and  Cohen \cite{YangCohenmori2012,YangWT2011PRL,CohenMoriYang2008} that for that reason DFAs should be favored for which the LUMO orbital energy would be $-A$ since
 \begin{equation}\label{eq:dEdnL}
 \partial E^{DFA}/\partial n_{L} = \epsilon_L=-A
 \end{equation}
The approximate functional (DFA) should then obey the Slater relation $\partial E^{DFA}/\partial n_{i} = \epsilon_i$.  The Slater relation \cite{SlaterMann1969,Slater1972} holds for many approximate functionals where occupation numbers of the orbitals have been introduced in a specific way (but not for all such functionals \cite{Baerends2018JCP}). (Note that Slater's relation for such approximate $E^{DFA}$s does not suffer from the problem with the analogous Janak theorem of Kohn-Sham DFT exemplified with Eq.\ \eqref{eq:partialEpartialn_i}.) 
We should caution that the density change upon an infinitesimal increase of the density by $\delta n_L|\psi_L^N(\br)|^2$ does not obey the PPLB prescription that an infinitesimal density change should bring in $\rho_0^{N+1}$ density, see Eq.\ \eqref{eq:N+omega}:
\begin{align}
\delta \rho^{PPLB} &= -\delta\omega \rho_0^N +\delta\omega \rho_0^{N+1} \notag. \\
&= -\delta\omega\sum_i^N |\psi_i^N(\br)|^2 +\delta\omega \sum_j^{N+1} |\psi_j^{N+1}(\br)|^2
\end{align}
The orbitals $\{\psi_i^{N+1}\}$ of the $(N+1)$-electron system will typically all be more expanded and at higher orbital energies than the corresponding orbitals $\{\psi_i^N\}$  of the $N$-electron system. Because $\delta n_L|\psi_L^N(\br)|^2 \ne \delta \rho^{PPLB} $ one cannot conclude that adherence to the PPLB straight-lines picture for the energy requires relation \eqref{eq:dEdnL} to hold (and we have denied a physical basis for such a requirement anyway). Nevertheless, functionals for which the Slater relation holds and for which $\epsilon_L \approx -A$ have, from a pragmatic point of view, the advantage that the local approximation does not lead to poor results for dissociated systems with overall an additional electron (yielding fractional electron fragments), as conventional (semi-)local functionals used to do \cite{Bally1997}.  They have the disadvantage that the LUMO and higher virtual orbitals are then very high lying, close to the  energy zero, and therefore are unduly diffuse \cite{BaerendsGritsenkovanMeer2013,vanMeerGritsenkoBaerends2014b}. This is a well-known deficiency of the Hartree-Fock virtual orbitals which also have $\epsilon_L \approx -A$. It also has the disadvantage that then the HOMO-LUMO gap is not a good approximation of the first excitation energy. The exact KS model generates virtual orbitals that are much lower lying, for which the HOMO-LUMO gap does approximate the first excitation energy very well \cite{SavinUmrigarGonze1998,Dixon2000JCP,vanMeerGritsenkoBaerends2014b}. The exact Kohn-Sham virtual orbitals are not unduly diffuse but represent the excited electron very well, so that most excitations can be expressed as single (or a few) orbital-to-orbital transitions \cite{vanMeerGritsenkoBaerends2014b}. The more realistic orbital energy spectrum from accurate model KS potentials (compared to the poor potentials resulting from conventional local and hybrid functionals) greatly improves excitation energies, in particular the Rydberg and mixed valence-Rydberg transitions.  The local potentials resulting from the OEP procedure have orbital energies closer to the exact KS ones and therefore have similar advantages for excitation energy calculations \cite{DellaSalaGoerling2003IJQC}.
 
\subsection{Ensembles in DFT}\label{subsec:GOK}
In quantum mechanics an ensemble usually just means a mixture of single-state density matrices. This is something different than the Gibbsian ensembles in statistical mechanics. The quantum mechanical ensembles are uncontroversial and in fact play an important role in DFT. We mention two cases.\\
\\
In the first place an equi-ensemble of the ground state and an excited state has been introduced as a means of obtaining the excitation energy by Theophilou \cite{Theophilou1979}. This has been extended by Gross, Oliveira and Kohn to ensembles with more excited states \cite{GrossOliveiraKohn1988-I,GrossOliveiraKohn1988-II,OliveiraGrossKohn1988-III}. This ensemble approach for excitation energies is currently receiving considerable interest \cite{Senjean2018,Fromager2019,Burke2017,Gould2018}. 
We emphasize that this type of  ``ensemble'' is something very different from statistical mechanical ensembles,  like the ``canonical ensemble'' and ``grand canonical ensemble''. There is no connection with thermodynamics, in this application the density matrix and ensemble are just elements of the edifice of quantum mechanics. (The names ``density matrix'' and ``ensemble'' have of course originated from the link with statistical mechanics \cite{DiracPrinciples1930,Neumann1932,FeynmanHibbs1965}, but the concepts are now independent of this context.)  There is no theoretical problem with the ensemble approach to excitation energies. It is interesting to observe that Levy has used this ensemble formulation for excitation energies to investigate the relation between excitation energies and Kohn-Sham orbital energies \cite{Levy1995}. The first excitation energy, for instance, is not equal to the KS orbital energy difference between HOMO and LUMO, $\epsilon_L-\epsilon_H$, for an $N$-electron system. The LUMO must be raised by a constant 
\begin{equation}
\Delta v_{xc}= \lim_{w \to 0} \left. \frac{\partial E_{xc~}^w[\rho] }{ \partial w}\right|_{\rho=\rho_w}
\end{equation}
Here $E_{xc}^w$ is the exchange-correlation energy that would yield the correct energy for an ensemble $E_w=(1-w)E_{GS}+wE_{H \to L}$ from a KS calculation with fractional occupations of the HOMO and LUMO. This upshift by a constant upon admixing an infinitesimal amount of the excited state density is a genuine discontinuity, reflecting the discontinuous change of the density by admixing of the different excited state density to the ground state density, $\rho_w=(1-w)\rho_{GS} + w\rho_{H\to L}$. But its derivation does not rely on or need any result from the case of an ensemble of $N$- and $(N+1)$-electron densities. 
The constant in this case arises from the density and energy of an excited state being different from those of the ground state. The theory does not provide an estimate of the magnitude of the constant $\Delta v_{xc}$. Fortunately, the quantitative magnitude of the deviation of excitation energy from KS orbital energy difference is generally quite small, at least for accurate KS orbital energies \cite{SavinUmrigarGonze1998,vanMeerGritsenkoBaerends2014b} (this is not true in general for the orbital energies of most DFAs, notably not for the Rydberg orbital energies of DFAs \cite{vanMeerGritsenkoBaerends2014b}).\\
\\
A second occurrence of ensembles in DFT is in the case of non-pure-state $v$-representable ground state densities.  Levy \cite{Levy1982} and Lieb \cite{Lieb1983} proved that in case of degenerate ground states some ground state densities are only ensemble $v$-representable. This has acquired some practical importance when it was discovered that there are cases where the density of a nondegenerate ground state wavefunction can only be represented by an ensemble density of a degenerate KS ground state \cite{SchipperGritsenko1998,Schipper1999,UllrichKohn2001}. This appears to be  connected to strong (nondynamical) correlation. The strong mixing in that case of a few electron configurations in the wavefunction then leads in the KS system to the description of the density by an ensemble of KS states representing the mixing electron configurations. In that case the KS states (determinants) are degenerate, the HOMO being degenerate. The KS ensemble is then an example of Levy and Lieb's ground state ensemble, but now for the KS noninteracting electron system.\\
The practical relevance of this type of ensemble has been demonstrated in a series of papers by Filatov (see review \cite{Filatov2015}) who developed the spin-restricted ensemble-referenced Kohn-Sham methods  (REKS) precisely for the cases where the density is no longer pure-state $v_s$ representable in the Kohn-Sham system but is only ensemble $v_s$ representable. This applies to many cases including diradicaloids and excited states (conical intersections) \cite{FilatovShaik2000,KazaryanFilatov2008,Filatov2016}. \\
\\
It is unfortunate that the terminology ``ensemble DFT'' is now gaining traction, comprising on the one hand the PPLB approach with its derivative discontinuity and on the other hand the ensemble  approaches for excitation energies and for non-pure-state representable Kohn-Sham cases. We stress that these are very different theoretical constructs.

\section{Final remarks}\label{sec:Conclusions}
We have given arguments why the statement ``$\partial E/\partial N$ is the chemical potential of the electrons in the molecule'' is wrong. It is wrong on two counts: it invokes a quantity ($\partial E/\partial N$) that has no physical meaning for an atom or molecule, and uses a thermodynamical concept (chemical potential) that does not refer to any property of the electrons in an energy eigenstate of such a small system. Few-electron systems like atoms and molecules can lose or gain an electron, with corresponding energy changes giving the ionization energy $I$ and electron affinity $A$. There is not a third energetic characteristic of the electrons which could be called ``the chemical potential''. The quantity $\partial E/\partial N$ is not defined for systems like atoms and molecules. We have noted that the solution of the HK based  Euler-Lagrange variational equation \eqref{eq:Euler-Lagrange} requires that the energy $E_v[\rho]$ is extended into the nonphysical domain of noninteger electron number in such a way that the derivative $\partial E/\partial N$ exists. It must be continuous at the $N$ point, it should not have different left and right derivatives (be discontinuous). The actual magnitude of the (continuous) derivative is arbitrary (it depends on the chosen extension of $E_v[\rho]$). The suggestion that ``exact DFT'' would have linear energy behavior with derivative  $-A$ between $N$ and $N+1$  and with derivative $-I$
between $N$ and $N-1$  would imply that the Euler-Lagrange equation of DFT is anomalous, since the discontinuity of the derivative at the integer $N$ point would preclude determination of the Lagrange multiplier as $\partial E/\partial N$.\\
\\
The heart of the problem with $\partial E/\partial N$ is that no physical meaning can be given to systems with a fractional number of electrons, such as $(N+\omega)$. This also leads to the denial of such physical meaning to the linear energy picture of Fig.\ \ref{fig:straightlines} other than that $\overline{N}$ is the average electron number for two states of different electron numbers which constitute an ensemble with mixture parameter $\omega$, and $\overline{E}^{PPLB}$ the average energy. Such mixtures have also been considered with  probabilities patterned after those of the grand canonical ensemble of statistical mechanics \cite{Gyftopoulos1968,PerdewParrLevyBalduz1982}, see section \ref{sec:PPLB}.  In that case the  behavior of Fig.\  \ref{fig:EStraightlines} is obtained, with collapse of the $\overline{E}(\overline{N})$ curve to just the points $(Z',E_0(Z')$. Neither the straigt-lines picture of Fig.\ \ref{fig:straightlines} nor the dashed lines of Fig.\ \ref{fig:EStraightlines} represent physical behavior of an atom or molecule.  \\
\\
We have been discussing the  Euler-Lagrange equation \eqref{eq:Euler-Lagrange} and other issues which pertain to the theory (DFT, i.e.\ quantum mechanics) of electrons in atoms and molecules, where temperature does not play a role.   Elevated temperature effects can of course be described with statistical mechanics. For instance, at (very) high temperature a macroscopic gas of H atoms may exhibit ionization, meaning that an equilibrium is established in the gas between H atoms, free electrons, and H$^+$ ions, see Refs \cite{Reif1965,Hill1960,LandauLifshitz1980} and Appendix \ref{app:reservoir}. In this case we are dealing with thermodynamic systems, in principle macroscopic with a defined pressure for the gases of each type of particle (atoms, ions, electrons). These are traditional systems for the application of thermodynamics and statistical mechanics. Ignoring the population of electronically excited states (which however will be important at such high temperatures that ionization becomes measurable), one could describe the fraction of H$^+$ ions by way of fractional occupation of the $1s$ orbital. This does not mean of course that any H atom/ion would exist in the gas with a fractional number of electrons. Just the \textit{average} number of electrons on an H becomes fractional.\\
Fractional occupations are also well known as the Fermi-Dirac distribution of electrons over single particle states in free-electron models of an electron gas at elevated temperatures, cf.\ Eq.\ \eqref{eq:niFD}. This is again just a way of describing the distribution of such a system over ground state and excited states at nonzero $T$, see Ref.~\cite{AshcroftMermin1976}, Ch. 2.  An electron gas in a potential with a defined $\mu$ (so no band gap) is a model for metals. The generalization of DFT to include temperature dependence for such a system was proposed long ago by Mermin \cite{Mermin1965}. For a gas of electrons moving in an external potential at $T \ne 0$ he established the one-to-one correspondence of the external potential and the density. Mermin uses that his system of electrons has a defined temperature $T$ and chemical potential $\mu$, signalling that we are dealing with a thermodynamic system. Not only DFT, also other electronic structure theories may be extended to incorporate temperature effects in extended electronic systems where $T$ and $\mu$ are defined quantities. The pioneering work in 1963 of Mermin on Hartree-Fock for the electrons at finite $T$ \cite{Mermin1963} should be mentioned as well as the recent upsurge of interest, for instance the work by Hirata and coworkers on one-dimensional  solids at finite temperature, with both Hartree-Fock approximation and various correlated methods \cite{Hirata2014,Hirata2015}. See also recent work by Harsha et al.\ \cite{Harsha2019,Harsha2020} and White and Chan \cite{WhiteChan2018}. Finite temperature effects in extended systems with defined $T$ and $\mu$ are of course well known in many-body (perturbation) theories \cite{FetterWalecka1971,NegeleOrland1988,GrossRungeHeinonen1991,StefanucciLeeuwen2013}. This does not imply that thermodynamic properties would exist for a small finite-electron system like an atom or molecule.   \\
\\
Finally we have noted that ``ensemble DFT'' is not a good common denominator for on the one hand for instance the T-GOK ensemble approach to excitation energy calculations \cite{Theophilou1979,GrossOliveiraKohn1988-I,GrossOliveiraKohn1988-II,OliveiraGrossKohn1988-III,Levy1995,Senjean2018,Fromager2019,Burke2017,Gould2018}, and/or the occurrence of ensembles to describe densities of degenerate ground states \cite{Levy1982,Lieb1983,SchipperGritsenko1998,UllrichKohn2001}, as employed in the REKS method \cite{Filatov2015,Filatov2016}, and on the other hand the use of ensembles of ground states of different electron number \cite{PerdewParrLevyBalduz1982}.  While there is no objection to the former, we have warned against the pitfalls that open up when unwarranted conclusions are drawn from behavior of the latter.\\
\\
\textbf{Conflicts of Interest.}\\
There are no conflicts of interest to declare.\\
\\
\textbf{Acknowledgement}\\
I am grateful to Wim Briels for sharing with me his insights in statistical mechanics and thermodynamics, and to Kieron Burke for lively disputes on the subject matter of this paper.

\newpage

\appendix
\section{Elements of the statistical mechanical underpinning of thermodynamics}\label{app:statistics}
The electrons in a molecule do not constitute a macroscopic thermodynamic system for which the concept of a chemical potential and temperature for the particles comprising the system has been defined. 
However, PPLB assume that one may use results from statistical mechanics to treat such a system. There is now abundant reference in the DFT literature to  the chemical potential for the electrons in a molecule and to the exchange of these particles with the environment as being governed by the chemical potential, and to the grand canonical ensemble as providing a proper description. It is therefore appropriate to investigate the validity of these concepts for the electrons in a molecule. Many excellent textbooks give extensive expositions of thermodynamics and its underpinning by way of statistical mechanics  \cite{Tolman1938,Rushbrooke1949,Hill1956,Hill1960,Landsberg1961,Reif1965,LandauLifshitz1980,Pathria2011}. We briefly highlight a few points that are relevant here. Although unabashedly unoriginal, we need this exposition to establish the salient features of statistical mechanics upon which our criticism of the application of concepts from this theory to few-electron quantum mechanical systems (atoms and molecules) is based. \\
One important characteristic of a thermodynamic system is that it has to be macroscopic for the following reasons. For some concepts or derivations it is necessary that the thermodynamic limit can be reached, meaning that the particle number can be increased to, say, Avogadro's number (ca.\ $10^{23})$, keeping the density $N/V$ constant. It is also necessary that the temperature can be measured and that equilibrium exists in the sense that the temperature will be the same in different parts of the system.  Statements about the existence and measurement of a physical attribute called ``temperature'' as well as its transitivity and its role in defining equilibrium, feature in the literature as  the Zeroth Law of Thermodynamics \cite{FowlerGuggenheim1935}. The basis of the statistical mechanical derivation of the properties of a thermodynamic system is the realisation that the system, for which the macroscopic state is described with a few macroscopic variables (e.g.\ $N, V, T$) will in the course of time traverse an exceedingly large number of microstates which are all compatible with the macroscopic state but which differ in the states of the large number of particles comprising the system. When the movements of the particles are described classically this is simple: the position and momentum coordinates for all particles define a point $(q_1 ... q_N, p_1 ... p_n) \equiv (q, p)$ in phase space which travels along some path in phase space due to the constant changes in position and in momentum (e.g.\ due to collisions). When the particles are described quantum mechanically the assumption of constantly changing microstates is a bit more subtle. It would not be compatible with this fundamental viewpoint of statistical mechanics to assume  that the total macroscopic system can be in an energy eigenstate and therefore be stationary, not subject to change. Detailed arguments why this cannot be the case can be found in the cited literature, notably Tolman \cite{Tolman1938} who stresses that the principle of detailed balance also applies to a macroscopic system in quantum mechanics, and e.g.\ Hill \cite{Hill1956} and Landau and Lifshitz \cite{LandauLifshitz1980}.  Landau and Lifshitz \cite{LandauLifshitz1980} summarize this in the statement that it is impossible for a macroscopic system to be in an energy eigenstate due to the unavoidable disturbance by interaction with the outside world and the internal disturbances by density fluctuation and other perturbations. So it is generally accepted that also when quantum mechanics is applied the same assumption holds that the system traverses in the course of time the microstates compatible with the thermodynamic state of the system.    \\
Since the calculation of the time-dependent behavior of the system is out of the question, at least before computer simulations came around, a so-called representative ensemble is formed. The ensemble consists of very many mental copies of the system, each presenting the system in a particular microstate. Then the basic postulate of statistical mechanics asserts that there is no \textit{a priori} bias in the probability that the system be in some microstate: all microstates (or all points in phase space) compatible with the macroscopic state variables, are equally probable. The impossible task of calculating a desired property as a time-average of the system is now replaced by an ensemble average. Given the  equal probabilities for all microstates, this amounts basically, for a wanted property, to finding the probability that a microstate has a certain value for the desired property. Then an average over all the microstates can be taken. If the treatment is quantum mechanical, the notion that macroscopic systems cannot be in a stationary state, does not preclude the use of quantum mechanical states - either energy eigenstates or some set of other states compatible with the thermodynamic variables - as the microstates of the systems in the ensemble, if only their number is large enough and representative of the thermodynamic system.     \\
\\
The simplest example is the case of an assembly of $N$ independent classical particles within a (macroscopic) volume $V$ with total energy $E$ or within a narrow energy range $(E-\Delta E,E+\Delta E)$. The independent particles will have individual energies $\{\epsilon_i\}$. If there are $n_i$ particles having energy $\epsilon_i$  the constraints of fixed particle number $N$ and energy $E$ can be written
\begin{align}\label{eq:constraints}
(a) \sum_i n_i = N; \qquad (b) \sum_i n_i \epsilon_i =E 
\end{align}
If the particles are distinguishable (for instance by their positions in a crystal lattice if they are 3D harmonic oscillators as in the Einstein model for a crystal), the total number of configurations (``microstates'') for a particular distribution of occupation numbers $\{n_i\}$ is
\begin{equation}\label{eq:OmegaDist} 
\Omega(\{n_i\}) = \frac{N!}{n_1!n_2!...n_i!...}
\end{equation}
and the total number of microstates $\Omega$ is in principle obtained by summing over all sets of occupation numbers compatible with the constraints \eqref{eq:constraints}. We will try to determine the set of occupation numbers, indicated with stars, that give the largest term, $\Omega(\{n_i^\star\})$. It is always not just $\Omega$ but $\ln{\Omega}$ that is considered. This is computationally much more expedient, see below, and does not matter since $\ln{\Omega(\{n_i\})}$ and $\Omega(\{n_i\})$ have the maximum for the same same set of occupation numbers. The deeper reason is that $\ln{\Omega}$ of a macroscopic system is connected to the thermodynamic entropy of such a system,
\begin{equation}\label{eq:entropy}
S=k\ln{\Omega}. \\
\end{equation}
The equation $S=k\ln{\Omega}$  is the first and most important bridge from statistical mechanics to thermodynamics. \\
The well known results of statistical mechanics assert that, as a consequence of the huge number of particles and very large $\Omega(\{n_i\})$ of thermodynamic systems, and the fact that it is the logarithm of $\Omega$ that enters the entropy, the contributions of all other terms $\Omega(\{n_i\})$ than just the maximum one, $\Omega(\{n_i^\star\})$, make a negligible contribution to  the entropy $k\ln{\Omega}$.\\
When the total number of particles $N$ is very large, and the occupations $\{n_i\}$ as well, so that the Stirling approximations $\ln{N!}=N\ln{N}-N$ and $\ln{n_i!}=n_i\ln{n_i}-n_i$ can be used, $\ln{\Omega(\{n_i\})}$ reduces to
\begin{equation}
\ln{\Omega(\{n_i\})}=N\ln{N}-N -(\sum_i n_i \ln{n_i} -n_i).
\end{equation}         
The occupation numbers that  correspond to the maximum $\ln{\Omega(\{n_i\})}$ can be found by optimization with the Lagrangian
\begin{equation}\label{eq:Lagrangian}
\mathcal{L}=\ln{\Omega(\{n_i\})}-\alpha(N-\sum_i n_i) - \beta(E-\sum_i n_i \epsilon_i)
\end{equation}
The conditions  $\partial \mathcal{L} / \partial n_i =0$ lead immediately to the well known equations for the optimal occupation numbers
\begin{equation}\label{eq:nistar}
n_i^\star=e^{-\alpha} e^{-\beta \epsilon_i}
\end{equation}
 It can be demonstrated (by the Darwin-Fowler method \cite{Pathria2011}) that actually the average of the occupation numbers over all systems in a representative ensemble is equal to the calculated most probable occupations of Eq.\ \eqref{eq:nistar}. An ensemble that is representative of the thermodynamic system under study has very many (mental) copies of this system with each one in a specific microstate, chosen with equal probability for all microstates compatible with the thermodynamic state (in the present example determined by $(N,V,E)$, a so-called microcanonical ensemble). 
The Lagrange multipliers can in principle be solved from the constraints of Eq.\ \eqref{eq:constraints}. However, we want to make the connection with thermodynamics, i.e.\ to determine values of $\alpha$ and $\beta$ in terms of thermodynamic properties. \\
\\ 
Thermodynamics has provided a framework for macroscopic systems of particles, with the introduction of quantities such as the (internal) energy $(E)$, temperature $(T)$, entropy $(S)$, volume $(V)$, chemical potential $(\mu)$ and derived quantities such as Helmholtz free energy $F=E-TS$, enthalpy $H=E+pV$ and Gibbs free energy $H-TS$. A fundamental relation is (cf.\ the First Law of Thermodynamics) 
\begin{equation}\label{eq:FirstLaw}
dE=TdS-pdV+\mu dN
\end{equation}
This relates energy change $dE$ to heat flow ($dQ=TdS$) plus work done by or on the system (in case of only volume as external mechanical parameter, just $-pdV$) and change in particle number, $\mu dN$, each particle addition (removal) bringing increase (decrease) of the internal energy (note the extensivity property of the energy).\\
Now from Eq.\ \eqref{eq:FirstLaw} several relations follow, for instance 
\begin{equation}\label{eq:temperature}
\left( \frac{\partial S}{\partial E} \right)_{N,V} = \frac{1}{T}
\end{equation}
and 
\begin{equation}\label{eq:chemicalpotential} 
\mu=\left(\frac{\partial E}{\partial N}\right)_{V,S} = - T \left(\frac{\partial S}{\partial N}\right)_{V,E}
\end{equation}
Comparing these thermodynamic results with the relations obtained from statistical mechanics affords a connection of the statistical mechanical parameters $\alpha$ and $\beta$ with thermodynamic quantities.  By substituting $n_i^\star$ \eqref{eq:nistar} into $\Omega(\{n_i\})$  \eqref{eq:OmegaDist} and using $S=k\ln{\Omega}$ \eqref{eq:entropy}, one obtains 
\begin{equation}\label{eq:kbeta}
\left( \frac{\partial S}{\partial E} \right)_{N,V} = k\beta
\end{equation}
Hence, using  \eqref{eq:temperature} 
\begin{equation}\label{eq:beta}
\beta=\frac{1}{kT}
\end{equation}
giving for the probability that a particle has energy $\epsilon_i$ the well known expression
\begin{equation}\label{eq:canonicalprobability}
p_i = \frac{n_i^\star}{N} =  \frac{e^{-\epsilon_i/kT}}{\sum_i e^{-\epsilon_i/kT}}
\end{equation}
where $(\sum_i e^{-\beta\epsilon_i})$ is called the (one-particle) partition function.\\
To determine $\alpha$ we use the $n_i^\star$ of Eq.\ \eqref{eq:nistar} and then \eqref{eq:chemicalpotential} gives, together of course with $S=k\ln{\Omega}$ \eqref{eq:entropy},  the result
\begin{equation}\label{eq:alpha}
\alpha=\frac{-\mu}{kT}
\end{equation}  
So \textit{if} the system we are representing with the ensemble is a thermodynamic system, for which the temperature $T$ is a defined attribute, and the chemical potential $\mu$ as well, the Lagrange multiplier $\beta$ can be identified with $1/kT$ and $\alpha$ with $-\mu/kT$. If not, then of course not.\\
\\
In connection with this thermodynamic interpretation of $\alpha$ and $\beta$ a short remark on the concept of equilibrium is in order \cite{Tolman1938,Landsberg1961,Pathria2011}. Let us consider two systems $A_1$ and $A_2$ with macrostates $(N_1,V_1,E_1)$ and $(N_2,V_2,E_2)$ respectively. The corresponding numbers of microstates are $\Omega_1(N_1,V_1,E_1)$ and $\Omega_2(N_2,V_2,E_2)$ and the total energy is $E^{total}=E_1+E_2$. If we bring these systems into contact, so that energy can be exchanged but not particles (so they are separated by a heat conducting wall through which particles cannot pass) at any time $t$  the total system will have a number of microstates dependent on the energies at that moment 
\begin{equation}
\Omega^{total}=\Omega_1(E_1)\Omega_2(E_2)=\Omega_1(E_1)\Omega_2(E^{total}-E_1)
\end{equation}
Equilibrium means that a distribution of energy over the two systems will be reached at which the number of microstates $\Omega^{total}(E^{total},E_1)$ is maximum. For macroscopic systems with their very many microstates it can be inferred that when the energy distribution is such that the number of microstates is a maximum, this number of microstates will be overwhelmingly larger than for an even slight departure from this optimum distribution. So the system will spend virtually all its time at this optimum energy distribution, which is what we perceive as equilibrium.  We must have for the equilibrium $E_1$ and $E_2$
\begin{equation}
\frac{\partial \Omega^{total}}{\partial E_1}=\frac{\partial \Omega_1(E_1)}{\partial E_1}\Omega_2(E_2)+\Omega_1(E_1)\frac{\partial \Omega_2(E_2)}{\partial E_2}\frac{\partial E_2}{\partial E_1}
\end{equation} 
With $\partial E_2/\partial E_1=-1$ one obtains as condition for equilibrium
\begin{equation}
\frac{\partial \ln{\Omega_1(E_1)}}{\partial E_1} = \frac{\partial \ln{\Omega_2(E_2)}}{\partial E_2}
\end{equation}
With the established connection $S=k\ln{\Omega}$ \eqref{eq:entropy} and Eq.\ \eqref{eq:temperature} we note that equilibrium corresponds to $T_1=T_2$. This also holds when $\Omega_1$ simply applies to a part of a total system, and $\Omega_2$ to the rest of the system, with the same type of particles everywhere and the particle density everywhere the same $(N_1/V_1=N_2/V_2)$. An essential requirement for a thermodynamic system in equilibrium is that temperature can be measured and be the same in every (macroscopic) part (cf.\  the fundamental  Zeroth Law of Thermodynamics). Allowing also particles to be exchanged between the two (sub)systems, it can be seen that the same holds for the chemical potential, i.e.\ $\mu_1=\mu_2$ for equilibrium between two systems with the same type of particle, or for (macroscopic) subpartitions of a thermodynamic system.\\
\\ 
At this point we stress that the derivation of the occupation number distribution \eqref{eq:canonicalprobability} hinges on two conditions: a) the statistical mechanical derivation requires primarily that huge numbers of particles are involved; b) the introduction of physical meaning (temperature, chemical potential) for the constants in the statistical expressions requires that the target system to which the statistics is applied is a \textit{bona fide} thermodynamic system, i.e.\ the system is macroscopic (in the order of $10^{23}$ particles) and in equilibrium, with a uniform temperature and chemical potential.\\
\\
So it is essential that we are dealing with a very large total particle number, but other aspects of our example above are not important. For instance, for the more relevant case (even classically) of indistinguishable particles, as in the ideal gas of noninteracting particles,  $\Omega(\{n_i\})$ has to be divided by $N!$,
\begin{equation}\label{eq:OmegaIndist}
\Omega(\{n_i\})=\frac{1}{n_1! n_2! ...n_i!..}.
\end{equation}
This makes little change since the $N!$ factor leads to the constant  term $\ln{N!}$ in the Lagrangian \eqref{eq:Lagrangian} that does not have any effect in the equations $\partial \mathcal{L} / \partial n_i =0$. The circumstance that we may also suppose the occupation numbers $\{n_i\}$ to be large, which made the derivation above  especially simple, allowing the Stirling approximation of $\ln{n_i!}$ to be used, is often not fulfilled. An obvious example is a gas of independent electrons (fermions) where the occupation of a given quantum mechanical one-particle state (e.g. a translational  energy eigenstate) can only be 0 or 1. Then of course the derivation has to be adapted.  In that case the occupation number distribution takes the form
\begin{equation}\label{eq:nistarFD}
n_i^\star=\frac{1}{e^{\alpha + \beta \epsilon_i} + 1}
\end{equation}
Again this can be related to the temperature and chemical potential of the electron gas, given of course that it conforms to the requirements of a thermodynamic system (very many particles, in equilibrium with temperature $T$ and chemical potential $\mu$). Again the relations \eqref{eq:alpha} and \eqref{eq:beta} are obtained for $\alpha$ and $\beta$ and the well-known Fermi-
Dirac occupations result for the average occupations
\begin{equation}\label{eq:niFD}
n_i^\star=\frac{1}{e^{-( \epsilon_i-\mu)/kT} + 1}
\end{equation} 
\\
The earlier discussion has used the microcanonical ensemble. Sometimes, mostly for calculational expedience, it is easier to use another type of ensemble, the canonical and grand canonical ensembles being best known. In these cases the constraints on the total energy of the thermodynamic system (canonical ensemble) or on both the number of particles and the total energy (grand canonical ensemble) are no longer maintained.  This does not prohibit the applicability of the results to the thermodynamic system, even if that still has fixed particle number and energy.  If the average particle number $\overline{N}$ or both the average energy ($\overline{E}$) and particle number $(\overline{N})$ are equal to the corresponding quantities in the thermodynamic system, the results are applicable since the deviations from the average have virtually no weight.\\
In connection with the main text our interest is in the grand canonical ensemble \cite{Hill1956,Hill1960,Landsberg1961,Reif1965,LandauLifshitz1980,Pathria2011}. We note that in that case only the \textit{total} number of particles in the whole ensemble of $\mathcal{N}$ members, and the \textit{total} energy $\mathcal{E}$ of the ensemble are fixed, which are in terms of the averages per system just $\mathcal{N}\overline{N}$ and  $\mathcal{N}\overline{E}$.  Let $n_{i,s}$ denote the number of systems that have at any time $t$ $N_i$ particles and energy $E_s$. So we have the relations
\begin{align}\label{eq:sumsgrandcanonical}
\sum_{i,s} n_{i,s} = \mathcal{N} \notag \\
\sum_{i,s} n_{i,s} N_i =\mathcal{N}\overline{N} \\
\sum_{i,s} n_{i,s} E_s = \mathcal{N} \overline{E} \notag
\end{align}     
Note that we no longer use independent particles, but allow for interactions between them so that the system has a total energy $E_s$ that may not be a sum of single particle energies.  Any set of numbers $\{n_{i,s}\}$ represents one of the possible distributions of particles and energy among the members of the ensemble. The number of ways in which this distribution can be realized  is
\begin{equation}
\Omega[\{n_{i,s}\}]=\frac{\mathcal{N}!}{\prod_{i,s}n_{i,s}!}
\end{equation}
Again performing an optimization of the distribution numbers so that $\ln{\Omega[\{n_{i,s}\}]}$ is maximized, with the constraints on total number $\mathcal{N}$ of systems in the ensemble and total energy $\mathcal{N} \overline{E}=\mathcal{E}$ of Eq.\ \eqref{eq:sumsgrandcanonical} with Lagrange multipliers $\alpha$ and $\beta$ one arrives at
\begin{equation}\label{eq:grandcanonicalprobability}
p(N_i,E_s)=\frac{n_{i,s}^\star}{\mathcal{N}}=\frac{\exp{(-\alpha N_i - \beta E_s)}}{\sum_{i,s}  \exp{(-\alpha N_i - \beta E_s)}}
\end{equation}
It can be proved again that the maximum $n_{i,s}^\star$ is actually equal to the ensemble average, $\langle n_{i,s}\rangle =n_{i,s}^\star$. {We stress that the derivation relies on huge numbers, this time a huge number of systems in the ensemble. In principle the limit $\mathcal{N} \to \infty$ can be taken and also the $n_{i,s}$ can be taken to be very large, so Stirling's approximations $\ln{\mathcal{N}!}=\mathcal{N}\ln{\mathcal{N}}-\mathcal{N}$ and $\ln{n_{i,s}!}=n_{i,s}\ln{n_{i,s}}-n_{i,s}$ can be used. The large number of systems is crucial to make the optimum distribution important in the sense that any other (even slightly deviating) distribution has comparatively negligible occurrence.\\
This ensemble is a mental construct that serves to obtain the time-average of quantities by averaging over the members of the ensemble. In order to establish that it is representative of a thermodynamic system, with  number of particles $N$ equal to the ensemble average $\overline{N}$ and energy $E$ equal to the ensemble average $\overline{E}$, we have to give the Lagrange multipliers $\alpha$ and $\beta$ of Eq.\ \eqref{eq:grandcanonicalprobability} a physical meaning. The bridge from statistical mechanical quantities to thermodynamic ones is again the First Law \eqref{eq:FirstLaw}. Expressing $\alpha$ and $\beta$ in terms of thermodynamic properties of the target system is more involved than in the simple case of the microcanonical ensemble above, see e.g. Ref.\ \cite{Pathria2011}, Ch.\ 4.3. 
But \textit{if} the system which we represent with a grand canonical ensemble is a \textit{bona fide} equilibrium thermodynamic system, with a temperature $T$ and with a chemical potential $\mu$ for the particles, to which the First Law applies, one finds again the meanings    $\alpha=-\mu/kT$ \eqref{eq:alpha} and $\beta=1/kT$  \eqref{eq:beta}. As for any thermodynamic system, the numbers of particles $N$ must be very large in order to have well defined $\mu$ and $T$. We thus arrive at the well known expression for the distribution of the members of the grand canonical ensemble that is representative of a $(\mu,V,T)$ thermodynamic system,
\begin{equation}\label{eq:gcprobability}
p(N_i,E_s)=\frac{\exp{(\mu N_i -E_s)/kT}}{\sum_{i,s}  \exp{(\mu N_i - E_s)/kT}}
\end{equation}
The denominator is the partition function in the case of the grand canonical ensemble.\\
\\
 Summarizing, Eq.\ \eqref{eq:gcprobability}  is valid for an ensemble with very many members and for a target thermodynamic system in equilibrium at a temperature $T$ with very many particles at chemical potential $\mu$. The ensembles discussed in the main text, with a few members which consist of the electrons in an atom or molecule in specific energy eigenstates, fall far short of  the requirements to qualify as  grand canonical ensembles: the number of ensemble members $\mathcal{N}$ should be very large to enable the statistical derivation of \eqref{eq:grandcanonicalprobability} and the numbers of particles (electrons) in each system should be very large in order for them to constitute a thermodynamic system in equilibrium and provide \eqref{eq:gcprobability}.

\section{Chemical potential and temperature of particles in small (non-thermodynamic) systems which are subsystems of thermodynamic systems}\label{app:reservoir}
In statistical mechanics often contact of a thermodynamic system with a huge reservoir is imagined in order to fix properties like the temperature, or both temperature and chemical potential, of the system. The precise details of the reservoir are sometimes not important (for instance when it serves to establish the temperature by heat exchange) but sometimes they are (for instance when particle exchange between reservoir and system is supposed to take place). In the latter case the particles must be identical in reservoir and system. A straightforward realization is to contemplate partitioning a macroscopic thermodynamic system into $\mathcal{N}$ parts by inserting walls that are thought to be permeable for heat and particles. In a further abstraction the walls  may be supposed to be not physical but just mathematical planes that effect the imaginary partitioning \cite{Landsberg1961,LandauLifshitz1980}. Obviously then the number of particles in each partition (``thermodynamic system") will fluctuate, all the other partitions taking the role of the reservoir. Clearly, to have equilibrium we need equality of the temperature and chemical potential in the central partition and the rest of the system. The set of all partitions may also serve as a physical realization of the ensemble (in this case a grand canonical ensemble).\\
Let us stress that the GC ensemble is a ($\mu$, $V$, $T$) ensemble: the members of the ensemble should be characterized by a (common) chemical potential $\mu$ and temperature $T$, which is thought to be effected by embedding in a huge reservoir (which may or may not be formed by all the other members). The summation in the grand partition function extends in principle over all particle numbers, and the associated energy levels, and the constancy of $\mu$ and $T$ might seem a bit problematic at the very low particle numbers. However, this is a moot point,  the probability distribution \eqref{eq:grandcanonicalprobability} peaks extremely at the particle number and energy of the actual thermodynamic system and the few terms at low particle number have essentially zero contribution. \\
\\
The role of the reservoir to establish temperature and chemical potential is unproblematic when the system itself is a thermodynamic system in equlibrium (both within itself and with the reservoir) because then the chemical potential and temperature of the particles in the system are unambiguously defined. But what about a system that is so small that it does not qualify as thermodynamic system, and $\mu$ and $T$ cannot be defined for the isolated small system? That this is possible is the underlying assumption when one imagines an atom to be ``in contact'' with a reservoir, from which the electrons in the atom are supposed to derive a chemical potential and temperature. \\
This is a subtle issue. In what sense such thermodynamic attributes can be associated with small (non-thermodynamic) systems may be elucidated with two examples that are discussed below.\\

a) Thermal ionization of atoms.\\
Perhaps the closest we can come to the physical realization of the concept of electrons in atoms and their ions in contact with a reservoir that determines temperature and chemical potential is the case of  thermal ionization of atoms or molecules. This ionization can be realized in a mixed gas of electrons and M and M$^+$ particles in a given container at such high temperature that there is at least some degree of ionization. One can put the container in a reservoir (huge heat bath) to fix its temperature. If one considers a thought experiment where the (electrons in the) molecules are brought into contact with an electron reservoir, the gas of electrons in the container can be taken to be a physical realization of the reservoir with which the M and M$^+$ molecules are in contact. We assume that an equilibrium will be established in this gas of M atoms, M$^+$ ions and ``free'' electrons. There must be interaction between the particles, in this case pretty violent in order to lead to ionization, but this is allowed and even necessary  in statistical mechanics in order to attain and maintain equilibrium. Interaction of the reservoir with the system is always necessary, for the reservoir to exert its function of providing heat and particles, and for equilibrium to be established between reservoir and system. It is only required that the particles are ``free'' in the sense that they spend an overwhelmingly large part of their time as undisturbed free particles, with only brief moments of the disturbances that effect the equilibrium.
The equilibrium 
\begin{equation}\label{eq:e+M+reaction}
\text{M}^+ +e^- \leftrightarrows \text{M}
\end{equation}
is just like a chemical equilibrium between atoms A and B that can form a molecule AB:
\begin{equation}\label{eq:A+Breaction}
\text{A} + \text{B} \leftrightarrows  \text{AB}.
\end{equation}
Thermal ionization is actually a textbook example of application of the theory of chemical equilibrium. At the very high temperatures where ionization starts to play a role the various particles will approach the behavior of an ideal classical gas in the given volume at the given temperature. The ionization problem then reverts to the simplest case of chemical equilibrium, namely the one in an ideal gas of the participating species (see e.g.\ Rushbrooke \cite{Rushbrooke1949}, Ch. XI, XII).  There are two independent components, for instance M and $e^-$, that the experimenter can vary, the actual numbers of free M$^+$, $e^-$ and combined M species in the container are then determined by the chemical equilibrium. For clarity of presentation we use the notation A, B and AB for the M$^+$, $e^-$ and M respectively, which emphasizes that M is really a composite particle. The actually present  species of free A atoms, free B atoms and free AB molecules are denoted 1, 2 and 12 respectively (note there are two independent components, and three species). The chemical potential of a component can be worked out as
\begin{equation}\label{eq:mu_A}
\mu_A=\left(\frac{\partial F}{\partial N_A}\right)_{N_B,V,T} = -kT\ln{\frac{z_A}{N_1^\star}}
\end{equation}
 where $N_1^\star$ is the number of particles of species 1 (free M$^+$ ions in the container) at equilibrium and $z_A$ is the single particle partition function $z_A=\sum_i e^{-\epsilon^A_i/kT}$ (forgetting about possible degeneracies).  In this derivation full account has to be taken of the fact that the free energy has to be expressed with the partition functions of all species,
 \begin{equation}
 F=-kT\ln{ \frac{z_A^{N_1^\star} z_B^{N_2^\star} z_{AB}^{N_{12}^\star} }{N_1^\star ! N_2^\star ! N_{12}^\star !} } 
 \end{equation} 
 with the relations
 \begin{align}\label{eq:Ftotal}
 N_1^\star + N_{12}^\star = N_A \notag \\
 N_2^\star + N_{12}^\star =N_B
 \end{align}
It is remarkable that the chemical potential of A particles is just the same as the chemical potential $\mu_1$ of a gas of $N_1^\star$ independent A 
particles in the given volume at temperature $T$. But $\mu_A$ is a \textit{global} property. We can add A particles to the container and have to wait for equilibrium to be established before we know how much the number $N_1^\star$ is increased and how many A particles have been used to increase the number $N_{12}^\star$ of AB molecules. We can express this by saying that $\mu_A$ is a property of the A particles in the container that cannot be reserved for only the free A particles, or for (only) the A particles in AB molecules, or indeed to a single A particle. $\mu_A$ pertains to all A particles collectively. This is sometimes expressed by saying that also the A particles in the AB molecules ``have'' chemical potential $\mu_A$. Similarly,  the particles B (i.e.\ the electrons in our case) ``have'' the chemical potential $\mu_B$, which is equal to $\mu_2$ of a free electron gas with $N_2^\star$ particles in the given volume, but $\mu_B$ is again a property of the B particles (the electrons) collectively, including those in the AB particles  (i.e\ M in \eqref{eq:e+M+reaction}).  The chemical potentials  $\mu_A$ and $\mu_B$ are global properties. This concept of a global property is more readily understandable for the temperature: it is very clear that we cannot measure the temperature of a single molecule, only the speed of a molecule can be measured at a certain instant. Such a measurement tells nothing about the temperature of the macroscopic system. The temperature is an additional piece of information on the equilibrium distribution of the particles over the accessible energy states, and does not affect the energy of a state. This is how the statement: the electrons have temperature $T$ has to be understood. In the same way the global chemical potentials $\mu_A$ and $\mu_B$ of the A and B components (together with the chemical potentials $\mu_1$, $\mu_2$ and $\mu_{12}$ of the species that are present) tell about the equilibrium distribution of A and B particles over free A and free B and composite AB particles \cite{Rushbrooke1949}. This is expressed in the Law of Mass Action. It is to be noted that it is not possible to calculate in an independent way the chemical potential of the A particles in AB molecules. Therefore, although it would be allowed to say that at equilibrium the chemical potential of the A particles is ``the same'' in the gas and in the AB molecules, this is more semantics than an operational statement: one cannot determine equilibrium from the requirement that the chemical potential in the gas and in the AB molecules is the same, since the latter cannot be determined independently. These chemical potentials have no bearing on the energy eigenstates of the particles of the various species which remain determined by just the Schr\"{o}dinger equation. \\

b) Thermodynamics of small (sub)systems.\\
It is sometimes possible to obtain results for small subsystems that are not by themselves a thermodynamic system (being too small) by treating the subsystem with apparently statistical mechanics relations \cite{Brielscomment,Hill1960}. The prime example is a lattice of adsorption sites, to which molecules can be adsorbed.  The simplest case is $M$ sites that can adsorb one molecule and $N \ll M$ adsorbed molecules that can exchange position with another molecule or an empty site (there is an equilibration mechanism). When one considers this phase to be in equilibrium with a free electron gas of certain pressure (temperature $T$ and chemical potential $\mu$) the Langmuir adsorption isotherm can be derived. An extension would be to have $M$ sites that can adsorb up to a maximum $m$ of molecules (particles) per site, with ensuing site energies $E_j(N)$ for $N$ particles adsorbed to the site. The analogy with our case of nuclei of charge $Z$ that can ``adsorb'' up to a maximum of $Z+1$ electrons with ensuing energies $E_j(N), 0 \le N \le Z+1$, is evident. It should be emphasized that one starts here with a macroscopic system with very many sites, that are independent (noninteracting). Now again we assume this phase of adsorbed particles (also called a lattice gas) to be in equilibrium with a gas phase of the particles. Temperature $T$ and chemical potential $\mu$ are to be fixed by contact of this whole system with a suitable reservoir.\\
Considerable simplification in the treatment may now be achieved due to the independence of the $M$ sites. The partition function for a site with $N$ particles is defined as $z(N)=\sum_j e^{-E_j(N)/kT}$. It is possible to write the grand partition function for the lattice gas of adsorbed particles (in which a summation over all particle numbers from 0 to $mM$ is carried out) as a simple power of single-site ``grand partition functions'' (see Hill \cite{Hill1960}, Ch. 7.2):
\begin{equation}
Z^{GC}(\mu, T, M)=\xi(\mu,T)^M
\end{equation} 
with 
\begin{equation}\label{eq:xi}
\xi(\mu,T)=\sum_{N=0}^m z(N)e^{\mu N/kT}=\sum_{N=0}^m \sum_j e^{(\mu N-E_j(N))/kT}
\end{equation}  
$\xi(\mu,T)$ looks like a grand canonical partition function of a single site, i.e. of a small system of maximum $m$ particles with energies $E_j(N)$.  The important point, however, is that this is not a genuine  ``grand partition function''  for a macroscopic thermodynamic system with many particles, for which $\mu$ and $T$ are defined properties (can be measured). The maximum particle number $m$ is small. The quantities $\mu$ and $T$ featuring in $\xi(\mu,T)$ are determined by the true large thermodynamic system of which a single site is a subsystem. If such subsystems are independent of each other one can use the subsystem ``grand partition function''  $\xi(\mu,T)$ to simplify some calculations. For instance it can be shown that the average number of particles at a site can be determined from the site ``GC partition function''. To see this we consider a grand canonical ensemble of the lattice gas systems.  
The average number of particles in the lattice gas can now be found with the usual derivative of $\ln{Z^{GC}}$ with respect to $\mu$,
 \begin{align}\label{eq:Paverage}
\overline{P}&=kT\frac{\partial \ln{Z^{GC}(\mu,T)}}{\partial \mu}= \frac{kT}{Z^{GC}}M\xi^{M-1}\frac{\partial \xi}{\partial \mu} \notag \\
&= MkT\frac{\partial \ln{\xi}}{\partial\mu}
\end{align}
This shows that the average number of particles per site $\overline{N}$, which is $\overline{P}/M$, can also be obtained as
\begin{equation}
\overline{N}=kT\frac{\partial \ln{\xi}}{\partial \mu}
\end{equation}
which agrees with Eq.\ \eqref{eq:xi}.\\
So this leaves the impression that the determination of e.g.\ an average particle number can be achieved with a small system ``grand partition function''. But it is important to realize that the derivation proceeded from the total thermodynamic system, and that this system is needed to give meaning to the chemical potential $\mu$ and temperature $T$  that feature in the single site ``grand partition function''. These are collective properties of the macroscopic thermodynamic system of a gas of particles in equilibrium with an array of many adsorption sites. They cannot be determined in an independent way for the small system of maximum $m$ particles bound to a single site, they are simply not attributes of such a system.  And again, the energies $E_j(N)$ of the single site systems (solutions to the Schr\"{o}dinger equation) are input to the thermodynamic treatment, they do not depend on or are in any way affected by the $\mu$ and $T$ that feature in this application of statistical thermodynamics. 
 
\bibliographystyle{rsc}

\providecommand*{\mcitethebibliography}{\thebibliography}
\csname @ifundefined\endcsname{endmcitethebibliography}
{\let\endmcitethebibliography\endthebibliography}{}

\end{document}